\documentclass[a4paper,11pt]{article}

\usepackage[english]{babel}
\usepackage[latin1]{inputenc}
\usepackage[T1]{fontenc}
\usepackage{ae,aecompl} 
\usepackage{graphicx,color,setspace,amsmath}
\usepackage{amssymb}
\usepackage{natbib}
\usepackage{setspace}
\usepackage{mathrsfs}
\usepackage{lscape}
\usepackage{rotating}
\usepackage{verbatim}
\usepackage{amsfonts}  
\usepackage{amsbsy}    
\newcommand{\beq}{\begin{equation}}
\newcommand{\eeq}{\end{equation}}

\newcommand{\beqar}{\begin{eqnarray}}
\newcommand{\eeqar}{\end{eqnarray}}
\newcommand{\bit}{\begin{itemize}}
\newcommand{\eit}{\end{itemize}}
\newcommand{\benum}{\begin{enumerate}}
\newcommand{\eenum}{\end{enumerate}}
\newcommand{\barr}{\begin{array}}
\newcommand{\earr}{\end{array}}

\newcommand\eq[1]{(\ref{#1})}
\newcommand{\bfm}[1]{\mbox{\boldmath ${#1}$}}        

\newcommand{\jump}[2]{[\mbox{\hspace{-#1em}}[#2]\mbox{\hspace{-#1em}}]}

\newcommand{\bjump}[2]{\left[\mbox{\hspace{-#1em}}\left[#2\right]\mbox{\hspace{-#1em}}\right]}

\def\XXint#1#2#3{{\setbox0=\hbox{$#1{#2#3}{\int}$}
   \vcenter{\hbox{$#2#3$}}\kern-.5\wd0}}

\def\b0{\mbox{\boldmath $0$}}

\def\bb{\mbox{\boldmath $b$}}
\def\bc{\mbox{\boldmath $c$}}
\def\bd{\mbox{\boldmath $d$}}
\def\be{\mbox{\boldmath $e$}}

\def\bg{\mbox{\boldmath $g$}}

\def\bj{\mbox{\boldmath $j$}}

\def\bn{\mbox{\boldmath $n$}}

\def\bp{\mbox{\boldmath $p$}}
\def\bq{\mbox{\boldmath $q$}}

\def\bt{\mbox{\boldmath $t$}}
\def\bu{\mbox{\boldmath $u$}}

\def\bw{\mbox{\boldmath $w$}}
\def\bx{\mbox{\boldmath $x$}}
\def\by{\mbox{\boldmath $y$}}
\def\bz{\mbox{\boldmath $z$}}
\def\bA{\mbox{\boldmath $A$}}
\def\bB{\mbox{\boldmath $B$}}

\def\bE{\mbox{\boldmath $E$}}
\def\bF{\mbox{\boldmath $F$}}

\def\bI{\mbox{\boldmath $I$}}

\def\bM{\mbox{\boldmath $M$}}

\def\bQ{\mbox{\boldmath $Q$}}
\def\bR{\mbox{\boldmath $R$}}

\def\bU{\mbox{\boldmath $U$}}

\def\bW{\mbox{\boldmath $W$}}

\def\f0{\ensuremath{\mathbb{O}}}

\def\fR{\ensuremath{\mathbb{R}}}

\newcommand{\Gs}{\sigma}

\newcommand{\BGve}{\bfm\varepsilon}

\newcommand{\BGs}{\bfm\sigma}

\newcommand{\BGy}{\bfm\psi}

\newcommand{\mF}{\ensuremath{\mathcal{F}}}

\newcommand{\mJ}{\ensuremath{\mathcal{J}}}
\newcommand{\mK}{\ensuremath{\mathcal{K}}}

\newcommand{\mP}{\ensuremath{\mathcal{P}}}

\newcommand{\mS}{\ensuremath{\mathcal{S}}}

\newcommand{\bmA}{\mbox{\boldmath $\mathcal{A}$}}
\newcommand{\bmB}{\mbox{\boldmath $\mathcal{B}$}}

\newcommand{\bmD}{\mbox{\boldmath $\mathcal{D}$}}

\newcommand{\bmG}{\mbox{\boldmath $\mathcal{G}$}}

\def\Im{\mathop{\mathrm{Im}}}
\def\Re{\mathop{\mathrm{Re}}}

\newcommand{\sign}{\mathop{\mathrm{sign}}}

\begin{document}


\centerline{\Large \textbf{Boundary integral formulation for interfacial cracks}}

\centerline{\Large  \textbf{in thermodiffusive bimaterials}}

\begin{center}
L. Morini$\footnote{Corresponding author. Tel.: +39 0461 282583, email address: lorenzo.morini@unitn.it.}$ and A. Piccolroaz\\
\end{center}

\centerline{\emph{Department of Civil, Environmental and Mechanical Engineering, University of Trento,}}

\centerline{\emph{Via Mesiano 77, 38123, Trento, Italy.}}

\vspace{2cm}

\begin{abstract}

An original boundary integral formulation is proposed for the problem of a semi-infinite crack at the interface between two dissimilar elastic materials in the presence of heat flows
and mass diffusion. Symmetric and skew-symmetric weight function matrices are used together with a generalized Betti's reciprocity theorem in order to derive a system of integral equations that relate 
the applied loading, the temperature and mass concentration fields, the heat and mass fluxes on the fracture surfaces and the resulting crack opening. 
The obtained integral identities can have many relevant applications, such as for the modelling of
crack and damage processes at the interface between different components in electrochemical energy devices characterized by multi-layered structures (solid oxide fuel
cells and lithium ions batteries). \\


\end{abstract}

\newpage

\section{Introduction}
 Modelling of interfacial crack problems in elastic bimaterials in presence of thermodiffusion represents an important issue for many engineering
 applications. In particular, it is crucial for studying fracture initiation and propagation at the interface between different components of  
 electrochemical energy devices which are characterized by multi-layered structures, such as solid oxide fuel cells and lithium ions batteries. Indeed, due to the high operational
 temperature that can be reached and to the intense particle fluxes that are required for maintaining the electrical current, the components of these devices 
 are subject to severe thermomechanical stresses as well as stresses induced by the particle diffusion \citep{Anand1, Zhang1}. This
 can cause damage and crack formation compromising their performances in terms of power generation and
 energy conversion efficiency \citep{Lowrie1, Malz1, Gout1, Zhao1, Pharr1}. For this reason, the modelling of fracture processes at the interface between
 the components of such battery devices is fundamental to the prediction of these phenomena, and subsequently enables successful manufacture and 
 reliable performances of the systems. 
 
 Due to their distinctive feature of reducing by one the dimension of the considered problems,
 so that only the boundary or surface of the domain needs to be modeled, 
 boundary integral formulations are particularly suitable for studying multiphysics 
 phenomena such as dynamic and static fracture processes in thermodiffusive elastic materials.  
 Several boundary integral approaches have been proposed for solving crack problems in linear elastic, thermoelastic and thermodiffusive materials, 
 such as analytical techniques based on the method of singular integral equations \citep{Weaver1, BudRic1, LinkZub1}, 
 and numerical techniques based on the boundary element method \citep{RizzShip1, Brebbia1, Sladek1, Sladek2, Sladek3}. In all these approaches, 
 the displacements and stress fields are defined by integral relations involving the Green's functions,
 which need to be derived analytically in explicit form \citep{BigCap1} or computed numerically \citep{Ang1}.
 Altough Green's functions have been derived for several crack problems in linear thermoelastic and thermodiffusive elastic materials \citep{Sturla1, Hou1, Kumar1, Kumar2}, 
 their utilization for calculating physical displacements and stress fields on the
 crack faces requires challenging numerical estimation of integrals whose convergence should be assessed carefully. 
 Moreover, the approach based on the Green's function method works when the tractions and the thermal and diffusive stresses acting on the discontinuity surface
 are symmetric, but not in the case where asymmetric mechanical and thermodiffusive loading distributions are applied on the crack faces.
 
 In this paper, the problem of a semi-infinite quasi-static crack at the interface between two dissimilar thermodiffusive elastic materials is addressed by means of an
 origianl boundary integral formulation which avoids the use of the Green's functions and the challenging computations connected, without any assumptions regarding
 the symmetry of the loading and of the temperature and mass concentration profile at the interface. The general approach recently proposed in \cite{PiccMish3, MorPicc2, VelMish1} 
 and \cite{MishPicc2} for interfacial crack problems in isotropic and anisotropic elastic bimaterials, based on Betti's reciprocal theorem and weight functions theory, is extended
 in order to study fracture processes in presence of thermodiffusion. The volume integral terms present in the reciprocity identity, associated with the temperature
 and mass concentration effects \citep{Now1}, are converted into surface integrals through an exact transformation based on the notion of Lam\'e elastic potentials \citep{SlaughB} while
 assuming that the temperature and mass concentration are harmonic in the domain. The derived original form of 
 Betti's identity is used together with symmetric and skew-symmetric weight function matrices derived by \cite{PiccMish3} for formulating the considered 
 crack problem in thermodiffusive bimaterials in terms of singular integral equations. 
 
 The article is organized as follows: in Section \ref{prel}, the static governing equations for a linear elastic thermodiffusive media are formulated.
 Reciprocity indetities are introduced and the volume integral terms associated with the temperature and mass concentration fields are transformed into
 boundary surface integrals by introducing Lam\'e elastic potentials and applying second Green's theorem. In Section \ref{intercrack}, the problem of a
 quasi-static crack at the interface bewteen two dissimilar elastic thermodiffusive media is introduced. The weight functions, defined as a special 
 singular solution of the homogeneous traction-free problem are used together with the obtained Betti's identity for formulating the problem in
 terms of boundary integral equations. The case of a plane strain crack is analysed in Section \ref{ident}. Lam\'e potentials in the weight functions
 space are derived in closed form, and explicit integral identities relating the applied mechanical 
 loading, the profiles of temperature, mass concentration, heat and mass fluxes on the interface and the resulting crack opening are obtained. Finally,
 in Section \ref{examples}, the integral identities are used to study some illustrative examples of plane crack problems in the presence of thermodiffusion.
 Exact expressions for the crack opening and tractions ahead of the crack tip associated with the introduced temperature and heat flux distributions are
 derived, and the corresponding stress intensity factors are calculated in closed form.
 
\section{Preliminary results: governing equations and reciprocity theorem}
\label{prel}
In this Section, constitutive relations and static balance equation for linear thermodiffusive elastic media in infinitesimal deformations are introduced. 
Betti's integral identities are derived by means of the theorem of the reciprocity of work, and volume integral terms associated with heat flows and mass diffusion are
converted into surface integrals using an exact transformation based on the introduction of Lam\'e elastic potentials.
\subsection{Governing equations}
For a linear isotropic elastic body where temperature changes and mass diffusion are considered, the constitutive relationship between the stress $\BGs$ and the strain $\BGve$ is
given by \citep{Now1}:
\beq
\label{hooke}
\BGs=2\mu\BGve+\left(\lambda\mbox{tr}\BGve-\gamma_t\theta-\gamma_c\chi\right)\bI,
\eeq
where $\lambda$ and $\mu$ are Lam\'e's constants, $\theta=T-T_0$ is the temperature of the medium with respect to the reference state ($\BGve=0,\ T=T_0 \ C=C_0)$,
$\chi=C-C_0$ is the concentration of diffusing particles with respect to the natural state, $\bI$ is the identity matrix, $\gamma_t=(3\lambda+2\mu)\alpha_t$ and 
$\gamma_c=(3\lambda+2\mu)\alpha_c$ with $\alpha_t$ and $\alpha_c$ coefficients of thermal and diffusive expansions, respectively. Then in the natural state of the system we have 
\beq
\label{natural}
\BGve=0,\quad \theta=0, \quad \chi=0,
\eeq
and in order to describe the material using linear thermoelastic diffusion theory \citep{Now1,SherHam1} the values of $\theta$ and $\chi$ are assumed such that $|\theta/T_0|\ll1$ and $|\chi/C_0|\ll1$. 

In the static case, the equations of equilibrium are given by
\beq
\label{equilibrium}
\nabla\cdot\BGs+\bb=0,
\eeq
where $\bb$ represents the body forces. Substituting the constitutive relation \eq{hooke} in \eq{equilibrium}, the Navier's equations in terms of displacements $\bu$ are obtained
\beq
\label{navier}
\mu\Delta\bu+(\lambda+\mu)\nabla(\nabla\cdot\bu)+\bb=\gamma_t\nabla\theta+\gamma_c\nabla\chi.
\eeq

The conservation of energy and mass in steady-state conditions leads to 
\beq
\label{fluxeq}
\nabla\cdot\bq=W, \quad \nabla\cdot\bj=0,
\eeq
where $W$ is the heat generated for unit of time and volume by internal sources. For the considered linear isotropic thermodiffusive media, 
the heat flux $\bq$ and particles current $\bj$ can be expressed as 
\beq
\label{flux}
\bq=-k_t \nabla\theta, \quad \bj=-D_c \nabla\chi,
\eeq
where $k_t$ is the thermal conductivity coefficient and $D_c$ is the mass diffusivity of the material. Using relations \eq{flux} in \eq{fluxeq}, the following 
Poisson's and Laplace's equations for the steady-state temperature and mass concentration fields are derived:
\beq
\label{poisson}
k_t\Delta\theta+W=0, \quad \Delta\chi=0.
\eeq

\subsection{Reciprocity theorem: transformation of the volume integrals}
\label{RecSec}
Let us consider a thermodiffusive elastic body of volume $V$ subject to the action of body forces $\bb^{(1)}$, tractions $\bt^{(1)}$, heat sources $W^{(1)}$, surface heating to the 
temperature $\theta_b^{(1)}$ and mass concentration on the boundary  $\chi_b^{(1)}$. These causes are written symbolically in the compact form
\beq 
\mathcal{I}^{(1)}=\left\{\bb^{(1)}, \bt^{(1)}, W^{(1)}, \theta_b^{(1)}, \chi_b^{(1)}  \right\},
\eeq
and produce in the body the state characterized by displacements $\bu^{(1)}(\bx)$, temperature $\theta^{(1)}(\bx)$ and mass concentration $\chi^{(1)}(\bx)$:
\beq 
\mathcal{C}^{(1)}=\left\{\bu^{(1)}(\bx),\ \theta^{(1)}(\bx),\ \chi^{(1)}(\bx) \right\}, \quad \bx \in V.
\eeq
The stresses $\BGs^{(1)}$ and the strains $\BGve^{(1)}$ are connected by the relationship \eq{hooke}, where the dilatation is defined as  $\mbox{tr}\BGve=\nabla\cdot\bu$, and 
they are assumed to be continuous together with their first derivatives. The displacements, temperature and mass concentration are also continuous and moreover have continuous derivatives up to
the second order for $\bx \in V+\partial V$. These functions satisfy the fields equations \eq{equilibrium}, \eq{poisson}$_{(1)}$ and \eq{poisson}$_{(2)}$ with boundary conditions:
\beq
\bt^{(1)}(\bx)=\BGs^{(1)}(\bx)\bn(\bx), \quad \bx \in \partial V, 
\eeq
and
\beq
  \theta^{(1)}(\bx)=\theta_b^{(1)}(\bx), \quad \chi^{(1)}(\bx)=\chi_b^{(1)}(\bx), \quad \bx \in \partial V,
\eeq
where $\bn$ denotes the unit outward normal to the surface $\partial V$. 

Introducing another set of causes $\mathcal{I}^{(2)}$ and effects  $\mathcal{C}^{(2)}$:
\beq
 \mathcal{I}^{(2)}= \left\{\bb^{(2)}, Q^{(2)}, \bt^{(2)}, \theta_b^{(2)}, \chi_b^{(2)}\right\}, \quad \mathcal{C}^{(2)}=\left\{\bu^{(2)}(\bx),\ \theta^{(2)}(\bx),\ \chi^{(2)}(\bx) \right\},
\eeq
and applying the procedure illustrated in \citet{Now1, Now2} for elastic thermodiffusive media, the following reciprocity integral 
relations bewteen the two systems of causes and results are derived:
\begin{displaymath}
\int_{\partial V}\left(\BGs^{(1)}\bn\cdot\bu^{(2)}-\BGs^{(2)}\bn\cdot\bu^{(1)}\right)dS+\int_{V}\left(\bb^{(1)}\cdot\bu^{(2)}-\bb^{(2)}\cdot\bu^{(1)}\right)dV +
\end{displaymath}
\beq
\label{Rec1_mech}
+\gamma_t \int_{V}\left(\theta^{(1)}\nabla\cdot\bu^{(2)}-\theta^{(2)}\nabla\cdot\bu^{(1)}\right)dV+\gamma_c \int_{V}\left(\chi^{(1)}\nabla\cdot\bu^{(2)}-\chi^{(2)}\nabla\cdot\bu^{(1)}\right)dV=0,
\eeq
\beq
\label{Rec1_thetaa}
k_t\int_{\partial V}\left(\nabla\theta^{(1)}\theta^{(2)}-\nabla\theta^{(2)}\theta^{(1)}\right)\cdot\bn \ dS+\int_{V}\left(W^{(1)}\theta^{(2)}-W^{(2)}\theta^{(1)}\right)dV=0,
\eeq
\beq
\label{Rec1_chia}
\int_{\partial V}\left(\nabla\chi^{(1)}\chi^{(2)}-\nabla\chi^{(2)}\chi^{(1)}\right)\cdot\bn \ dS=0.
\eeq 

Expressions \eq{Rec1_mech} and \eq{Rec1_thetaa} have been extensively used in the literature in order to develop numerical buondary elements methods for studying both static 
and dynamic crack problems in elastic materials subject to thermal stresses \citep{Sladek1, Sladek2, Sladek3, DelleAli1}. In most of these approaches, starting from equations
\eq{Rec1_mech} and \eq{Rec1_thetaa}, integral expressions for the physical displacements, stresses and temperature $\bu^{(1)}, \BGs^{(1)}$ and $\theta^{(1)}$,  are derived as functions of 
the test quantities $\bu^{(2)}, \BGs^{(2)}$ and $\theta^{(2)}$, which in general are assumed to be fundamental solutions of the evolution equations 
or weight functions \citep{Brebbia1,RizzShip1}. Several analytical and numerical transformations have been proposed for reducing the volume integrals associated with body forces and coupling
between mechanical strains and temperature to surface integrals \citep{ ShiTan1, ShiTan2}. 

Here, we assume zero body forces and heat sources acting on the system, and we introduce an exact procedure for transforming coupling volume terms involving temperature 
and mass concentration in equation  \eq{Rec1_mech}. From Helmholtz's decomposition theorem, since both the displacements $\bu^{(1)}$, $\bu^{(2)}$ are solutions of the equilibrium equation \eq{navier},
they can be represented by \emph{scalar displacement potentials} $\varphi^{(1)}, \varphi^{(2)}$ and \emph{vector displacement potentials} $\BGy^{(1)}, \BGy^{(2)}$ \citep{SlaughB}:
\beq
\label{potentials}
\bu^{(1)}=\nabla\varphi^{(1)}+\nabla\times\BGy^{(1)}, \quad \bu^{(2)}=\nabla\varphi^{(2)}+\nabla\times\BGy^{(2)},
\eeq
substituting expressions \eq{potentials} into \eq{Rec1_mech}, the following equation is obtained:
\begin{displaymath}
\int_{\partial V}\left(\BGs^{(1)}\bn\cdot\bu^{(2)}-\BGs^{(2)}\bn\cdot\bu^{(1)}\right)dS+
\end{displaymath}
\beq
\label{Rec2_mech}
+\gamma_t \int_{V}\left(\theta^{(1)}\Delta\varphi^{(2)}-\theta^{(2)}\Delta\varphi^{(1)}\right)dV+\gamma_c \int_{V}\left(\chi^{(1)}\Delta\varphi^{(2)}-\chi^{(2)}\Delta\varphi^{(1)}\right)dV=0.
\eeq

The Green's second identity states that two arbitrary scalar functions, $\phi$ and $\vartheta$, must satisfy 
\beq
\int_{V}\left(\phi\Delta\vartheta-\vartheta\Delta\phi\right)dV=\int_{\partial V}\left(\phi\nabla\vartheta-\vartheta\nabla\phi\right)\cdot \bn \ dS.
\eeq
Substituting respectively $\theta^{(1)}, \theta^{(2)}, \chi^{(1)}, \chi^{(2)}$ to $\vartheta$ and $\varphi^{(1)}, \varphi^{(2)}$ to $\phi$, using equations \eq{poisson}
and remembering that zero heating sources are assumed, we get
\beq
\label{surf1}
\int_{V}\theta^{(1)}\Delta\varphi^{(2)}dV=\int_{\partial V}\left(\theta^{(1)}\nabla\varphi^{(2)}-\varphi^{(2)}\nabla\theta^{(1)}\right)\cdot \bn \ dS,
\eeq
\beq
\int_{V}\theta^{(2)}\Delta\varphi^{(1)}dV=\int_{\partial V}\left(\theta^{(2)}\nabla\varphi^{(1)}-\varphi^{(1)}\nabla\theta^{(2)}\right)\cdot \bn \ dS,
\eeq
\beq
\int_{V}\chi^{(1)}\Delta\varphi^{(2)}dV=\int_{\partial V}\left(\chi^{(1)}\nabla\varphi^{(2)}-\varphi^{(2)}\nabla\chi^{(1)}\right)\cdot \bn \ dS,
\eeq
\beq
\label{surf2}
\int_{V}\chi^{(2)}\Delta\varphi^{(1)}dV=\int_{\partial V}\left(\chi^{(2)}\nabla\varphi^{(1)}-\varphi^{(1)}\nabla\chi^{(2)}\right)\cdot \bn \ dS,
\eeq
substituting expressions \eq{surf1}-\eq{surf2} into \eq{Rec2_mech}, we finally derive
\begin{displaymath}
 \int_{\partial V}\left(\BGs^{(1)}\bn\cdot\bu^{(2)}-\BGs^{(2)}\bn\cdot\bu^{(1)}\right)dS+
\end{displaymath}
\begin{displaymath}
 +\gamma_t \int_{\partial V}\left(\theta^{(1)}\nabla\varphi^{(2)}-\varphi^{(2)}\nabla\theta^{(1)}-\theta^{(2)}\nabla\varphi^{(1)}+\varphi^{(1)}\nabla\theta^{(2)}\right)\cdot \bn \ dS +
\end{displaymath}
\beq
\label{Rec3a_mech}
+\gamma_c \int_{\partial V}\left(\chi^{(1)}\nabla\varphi^{(2)}-\varphi^{(2)}\nabla\chi^{(1)}-\chi^{(2)}\nabla\varphi^{(1)}+\varphi^{(1)}\nabla\chi^{(2)}\right)\cdot \bn \ dS = 0.
\eeq

By means of the proposed general procedure, the volume integral terms associated with thermal and diffusive stresses in the reciprocity identity \eq{Rec2_mech} has been reduced to surface integrals.
Remembering the fluxes definitions \eq{flux}, expression \eq{Rec3a_mech} can be written as follows
\begin{displaymath}
 \int_{\partial V}\left(\BGs^{(1)}\bn\cdot\bu^{(2)}-\BGs^{(2)}\bn\cdot\bu^{(1)}\right)dS+
\end{displaymath}
\begin{displaymath}
 +\gamma_t \int_{\partial V}\left(\theta^{(1)}\nabla\varphi^{(2)}-\theta^{(2)}\nabla\varphi^{(1)}\right)\cdot \bn \ dS + 
 \beta_t \int_{\partial V}\left(\varphi^{(2)}\bq^{(1)}-\varphi^{(1)}\bq^{(2)}\right)\cdot \bn \ dS +
\end{displaymath}
\beq
\label{Rec3_mech}
+\gamma_c \int_{\partial V}\left(\chi^{(1)}\nabla\varphi^{(2)}-\chi^{(2)}\nabla\varphi^{(1)}\right)\cdot \bn \ dS+
\beta_c \int_{\partial V}\left(\varphi^{(2)}\bj^{(1)}-\varphi^{(1)}\bj^{(2)}\right)\cdot \bn \ dS = 0,
\eeq
where $\beta_t=\gamma_t/k_t$ and $\beta_c=\gamma_t/D_c$. Assuming $W=0$, reciprocity identities \eq{Rec1_thetaa} and \eq{Rec1_chia} can also be expressed in terms of fluxes:
\beq
\label{Rec1_theta}
\int_{\partial V}\left(\bq^{(2)}\theta^{(1)}-\bq^{(1)}\theta^{(1)}\right)\cdot\bn \ dS=0, \quad \int_{\partial V}\left(\bj^{(2)}\chi^{(1)}-\bj^{(1)}\chi^{(2)}\right)\cdot\bn \ dS=0.
\eeq

In the cases where it is possible to determine explicitly the elastic potentials \eq{potentials}, using expression \eq{Rec3_mech} together with \eq{Rec1_theta} boundary integral 
formulation of static crack problems in thermodiffusive solids can be obtained avoiding numerical estimation of volume integrals \citep{ChengChen1}. In the next Sections, the obtained relation 
\eq{Rec3_mech} will be extensively applied in order to derive integral identities for the modelling of fracture phenomena at the interface between dissimilar
elastic materials in presence of thermodiffusion.
\begin{figure}[!htcb]
\centering
\includegraphics[width=9cm]{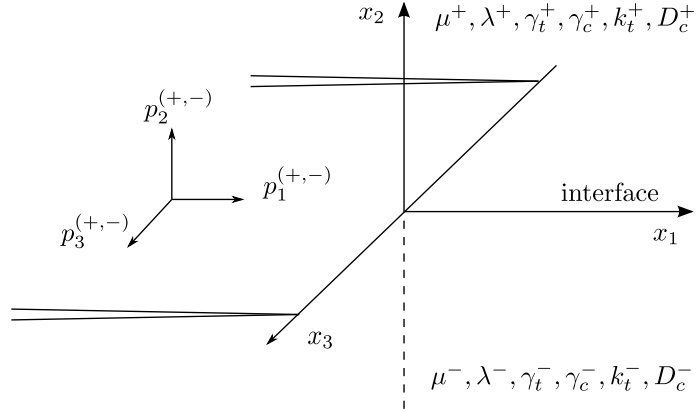}
\caption{\footnotesize Geometry of the model.}
\label{SOFCcrack}
\end{figure}

\section{Interfacial cracks in thermodiffusive media}
\label{intercrack}
A static semi-infinite crack between two dissimilar elastic materials in presence of thermodiffusion is condidered, the geometry of the system 
is shown in Fig.\ref{SOFCcrack}. No body forces, heat and mass sources are assumed. The crack is situated in the half-plane $\mathbb{R}^{2}_{-}=\left\{\bx=(x_1,x_2,x_3)\in 
\mathbb{R}^{3} : x_{1} < 0, x_2=0\right\}$, and general non-symmetric 
loading applied to the crack faces is assumed. Further in the text, we will use the superscripts $^{+}$ and $^{-}$ to denote the quantities related to the upper and the 
lower elastic thermodiffusive half-spaces, respectively. The applied loading can be decomposed in the symmetrical and skew-symmetrical parts, which are defined as follows \citep{PiccMish1}:
\beq
\label{load}
\left\langle \bp\right\rangle = \frac{1}{2}\left(\bp^{+} + \bp^{-}\right),\quad \jump{0.1}{\bp} = \bp^{+} - \bp^{-},
\eeq
where we used standard notations to denote the average, $\langle f \rangle$, and the jump, $\jump{0.1}{f}$, of a function $f$ across the plane containing the crack, $x_2=0$,
\beq
\left\langle f \right\rangle(x_1,x_3)=\frac{1}{2}[f(x_1,0^+,x_3)+f(x_1,0^-,x_3)], \quad  \jump{0.1}{f}(x_1,x_3)=f(x_1,0^+,x_3)-f(x_1,0^-,x_3).
\eeq

In the proposed model, thermal conduction and diffusion are both assumed to be isotropic, and then only normal stresses are induced by temperature changes and mass diffusion 
(see constitutive relation \eq{hooke}). 

Further in the paper, using the Betti identities \eq{Rec3_mech} and \eq{Rec1_theta}, the considered crack problem will 
be formulated in terms of boundary integral equations. In order to correctly apply expressions \eq{Rec3_mech} and \eq{Rec1_theta}, a preliminary discussion regarding the behaviour 
of the temperature and mass concentration functions in the upper and lower elastic thermodiffusive half-spaces is reported in next Section \ref{thermohalf}, and boundary conditions 
for these fields and the fluxes \eq{flux} are introduced.

\subsection{Temperature and mass concentration in thermodiffusive half-spaces}
\label{thermohalf}
Since zero heat and mass sources are assumed, the temperature $\theta$ and mass concentration fields are harmonic functions determined by the solution of a Laplace's equation assuming
the form of \eq{poisson}$_{(2)}$. Considering the upper elastic thermodiffusive half-space these equations become
\beq
\Delta\theta^{+}=0, \quad \Delta\chi^{+}=0, \quad x_2>0.
\label{harm1}
\eeq
Note that the conditions of harmonicity of the temperature and concentration fields \eq{harm1} must be satisfied in order to transform the volume integrals involved in
the reciprocity identity \eq{Rec2_mech} in surface integrals following the procedure explained in Section \ref{RecSec}.

The values of $\theta$ and $\chi$ for $x_2=0^{+}$, corresponding to the plane containing both the bounded interface and the crack,
are given by:
\beq
 \theta^+(x_2=0^+)=\theta(x_1,0^+,x_3), \quad \chi^+(x_2=0^+)=\chi(x_1,0^+,x_3).
\label{bound}
 \eeq
Replacing $\theta^{+}$ with $\theta^{-}$ and $\chi^{+}$ with $\chi^{-}$ a couple of Laplace's  equations identical to the \eq{harm1} and defined for $x_2<0$ 
are obtained for the lower thermodiffusive half-space, and then for $x_2=0^{-}$ boundary conditions analgous to \eq{bound} can be written.

The average and the jump of the temperature and concentration are defined on the whole plane $x_2=0$:
\begin{displaymath}
\left\langle \theta\right\rangle(x_1,x_3) =\frac{1}{2}[\theta(x_1,0^+,x_3)+\theta(x_1,0^-,x_3)], \quad -\infty<x_1, x_3<+\infty,
\end{displaymath}
\beq
\left\langle \chi\right\rangle(x_1,x_3) =\frac{1}{2}[\chi(x_1,0^+,x_3)+\chi(x_1,0^-,x_3)], \quad -\infty<x_1, x_3<+\infty,
\label{average_td}
\eeq
\begin{displaymath}
\jump{0.1}{\theta}(x_1,x_3)=\theta(x_1,0^+,x_3)-\theta(x_1,0^-,x_3), \quad -\infty<x_1, x_3<+\infty,
\end{displaymath}
\beq
\jump{0.1}{\chi}(x_1,x_3)=\chi(x_1,0^+,x_3)-\chi(x_1,0^-,x_3), \quad -\infty<x_1, x_3<+\infty.
\label{jump_td}
\eeq
Further in the text, we will refer to \eq{average_td} and \eq{jump_td} as to imperfect thermodiffusive interface conditions.
The values of the components of the fluxes normal to the interface plane, $q_2=\bq\cdot\be_2$ and $j_2=\bj\cdot\be_2$, for $x_2=0^{+}$ are
\beq
 q_2^+(x_2=0^+)=-\left.k_t^+\frac{\partial\theta^+}{\partial x_2}\right|_{x_2=0^+}=q_2(x_1,0^+,x_3), \quad 
 j_2^+(x_2=0^+)=-\left.D_c^+\frac{\partial\chi^+}{\partial x_2}\right|_{x_2=0^+}=j_2(x_1,0^+,x_3), \
\label{qj_int}
 \eeq
and similarly to the cases of temperature and mass concentration, expressions analogous to the \eq{qj_int} can be defined for the normal fluxes in the limit $x_2=0^{-}$.

The average and the jump of of $q_2$ and $j_2$ are also defined on the whole plane $x_2=0$:
\begin{displaymath}
\left\langle q_2 \right\rangle(x_1,x_3) =\frac{1}{2}[q_2(x_1,0^+,x_3)+q_2(x_1,0^-,x_3)], \quad -\infty<x_1, x_3<+\infty,
\end{displaymath}
\beq
\left\langle j_2 \right\rangle(x_1,x_3) =\frac{1}{2}[j_2(x_1,0^+,x_3)+j_2(x_1,0^-,x_3)], \quad -\infty<x_1, x_3<+\infty,
\label{average_flux}
\eeq
\begin{displaymath}
\jump{0.1}{q_2}(x_1,x_3)=q_2(x_1,0^+,x_3)-q_2(x_1,0^-,x_3), \quad -\infty<x_1, x_3<+\infty,
\end{displaymath}
\beq
\jump{0.1}{j_2}(x_1,x_3)=j_2(x_1,0^+,x_3)-j_2(x_1,0^-,x_3), \quad -\infty<x_1, x_3<+\infty.
\label{jump_flux}
\eeq

Both in the upper and lower half spaces, the temperatures and mass concentration fields are assumed to vanish at the infinity:
\begin{displaymath}
\theta^{+}=\chi^{+}=0, \ \mbox{for} \ x_1,x_3\rightarrow \pm\infty, \quad \theta^{+}=\chi^{+}=0, \mbox{for} \  x_2\rightarrow +\infty,
\end{displaymath}
\beq
\theta^{-}=\chi^{-}=0, \ \mbox{for} \ x_1,x_3\rightarrow \pm\infty, \quad \theta^{-}=\chi^{-}=0, \mbox{for} \  x_2\rightarrow -\infty.
\label{fields_inf}
\eeq

Since the heat and mass conduction problems \eq{harm1} are stationary, the flux functions $q_2$ and $j_2$ must satisfy the self-balance conditions on the boundary.
For the considered upper thermodiffusive half-plane this means that: 
\beq
\iint_{-\infty}^{+\infty}q_2(x_1,0^+,x_3)dx_1dx_3=0, \quad \iint_{-\infty}^{+\infty}j_2(x_1,0^+,x_3)dx_1dx_3=0, 
\label{balx1}
\eeq
where the definitions \eq{qj_int} for the values of the fluxes on the interface plane $x=0$ have been utilized. Note that the self-balance conditions
for the fluxes on $x_2=0^-$ are given by relations similar to the \eq{balx1}. Remembering the 
definitions \eq{average_flux} and \eq{jump_flux}, integral balance conditions analogues to the \eq{balx1} can be derived for the average and jump of the fluxes across the plane $x_2=0$.

\subsection{Boundary integral equations}
Considering the geometry of the model shown in Fig. \ref{SOFCcrack}, the Betti identities \eq{Rec3_mech} and \eq{Rec1_theta} are applied 
to a semi-spherical domain of radius $r$ in the upper and in the lower half spaces $\mathbb{R}^{3}_{+}$ and $\mathbb{R}^{3}_{-}$. 
In the limit $r\rightarrow\infty$, assuming that both the displacement fields $\bu^{(1)}$ and $\bu^{(2)}$ decay suitably fast at infinity and remembering
conditions \eq{fields_inf} for the temperature and mass concentration fields, the reciprocity relation \eq{Rec3_mech} in the upper half-space becomes:
\begin{displaymath}
 \int_{(x_2=0^+)}\left[\BGs_2^{(1)}(x_1,0^+,x_3)\cdot\bu^{(2)}(x_1,0^+,x_3)-\BGs_2^{(2)}(x_1,0^+,x_3)\cdot\bu^{(1)}(x_1,0^+,x_3)\right] \ dx_1 dx_3 + 
\end{displaymath}
\begin{displaymath}
 +\gamma_t^+ \int_{(x_2=0^+)} \left[ \theta^{(1)}(x_1,0^+,x_3)\frac{\partial \varphi^{(2)}(x_1,0^+,x_3)}{\partial x_2}-\theta^{(2)}(x_1,0^+,x_3)
 \frac{\partial \varphi^{(1)}(x_1,0^+,x_3)}{\partial x_2} \right] \ dx_1dx_3 +
\end{displaymath}
\begin{displaymath}
 +\beta_t^+ \int_{(x_2=0^+)}\left[ \varphi^{(2)}(x_1,0^+,x_3)q_2^{(1)}(x_1,0^+,x_3)-\varphi^{(1)}(x_1,0^+,x_3)q_2^{(2)}(x_1,0^+,x_3)  \right] \ dx_1dx_3 +
\end{displaymath}
\begin{displaymath}
 +\gamma_c^+ \int_{(x_2=0^+)} \left[ \chi^{(1)}(x_1,0^+,x_3)\frac{\partial \varphi^{(2)}(x_1,0^+,x_3)}{\partial x_2}-\chi^{(2)}(x_1,0^+,x_3)
 \frac{\partial \varphi^{(1)}(x_1,0^+,x_3)}{\partial x_2}\right] \ dx_1dx_3 +
\end{displaymath}
\beq
\label{biq_mech+}
 +\beta_c^+ \int_{(x_2=0^+)}\left[ \varphi^{(2)}(x_1,0^+,x_3)j_2^{(1)}(x_1,0^+,x_3)-\varphi^{(1)}(x_1,0^+,x_3)j_2^{(2)}(x_1,0^+,x_3)  \right] \ dx_1dx_3 = 0,
\eeq
where $\bu=[u_1,u_2, u_3]^T$ and $\BGs_2$ denotes the traction vector acting on the plane $x_2=0$: $\BGs_2=\BGs\be_2$. In the same limit $r\rightarrow\infty$, expressions 
\eq{Rec1_theta} assume the following form: 
\beq
\label{biq_term+}
\int_{(x_2=0^+)}\left[\theta^{(1)}(x_1,0^+,x_3)q_2^{(1)}(x_1,0^+,x_3)-\theta^{(2)}(x_1,0^+,x_3)
q_2^{(2)}(x_1,0^+,x_3)\right] \ dx_1dx_3=0,
\eeq
\beq
\label{biq_diff+}
\int_{(x_2=0^+)}\left[\chi^{(1)}(x_1,0^+,x_3)j^{(2)}_2(x_1,0^+,x_3)-\chi^{(2)}(x_1,0^+,x_3)j^{(1)}_2(x_1,0^+,x_3)\right] \  dx_1dx_3=0.
\eeq

 Considering the boundary $x=0^-$ instead of $x=0^+$, three similar expressions are derived for the lower half-space. In the cases where elastic potentials can be derived explicitly, 
 integral equations \eq{biq_mech+}, \eq{biq_term+} and \eq{biq_diff+} can be used together with their analogues in the lower half-plane in order to obtain expressions for the physical
 fields associated with the crack problem $\bu^{(1)}, \BGs_2^{(1)}, \theta^{(1)}, q_2^{(1)}, \chi^{(1)}$ and $j_2^{(1)}$. These physical quantities are evaluated by means of a set of
 predetermined auxiliary functions  $\bu^{(2)}, \BGs_2^{(2)}, \theta^{(2)}, q_2^{(2)}, \chi^{(2)}$ and $j_2^{(2)}$ satisfying the balance equations.
\subsection{Betti formula and weight functions}
 Following the approach proposed in several boundary element formulations of thermoelasticity \citep{ChengChen1, ShiTan2}, for the auxiliary solutions system we assume $\theta^{(2)}=q_2^{(2)}=0$
 and $\chi^{(2)}=j_2^{(2)}=0$. The displacement field $\bu^{(2)}$ is represented by the non-trivial singular solution of the homogeneous traction-free crack problem, commonly known as the
 weight function \citep{Bueck1, Bueck2}, and defined as follows
\beq
\label{weight}
\bu^{(2)}(x_1,x_2,x_3)=\bR\bU(-x_1,x_2,-x_3), 
\eeq
where $\bR$ is the rotation matrix: 
$$
\bR = 
\left[
\begin{array}{ccc}
-1 & 0 & 0 \\  0 & 1 & 0 \\ 0 & 0 & -1
\end{array}
\right].
$$

The transformation \eq{weight} corresponds to a change of coordinates consisting in a rotation of an angle $\pi$ around the $x_2-$axis, and it is connected to the fact that the weight function 
$\bU$ is defined in a different domain with respect to physical displacement, where the crack is placed along the positive semi-plane $x_1=0, x_2>0$ \citep{WilMov1}. 
The stress tension $\bfm{\varSigma}$ associated with the displacements $\bU$ is introduced:
\beq
\label{Sigma}
\BGs^{(2)}(x_1,x_2,x_3)=\bR \bfm{\varSigma}(-x_1, x_2,-x_3) \bR.
\eeq
Remembering the general definition \eq{potentials} and taking into account the transformation \eq{weight}, the following elastic potentials are introduced in the space of the weight functions:
\beq
\label{PhiPsi}
\varphi^{(2)}(x_1,x_2,x_3)=\varPhi(-x_1,x_2,-x_3), \quad \bfm{\psi}^{(2)}(x_1,x_2,x_3)=\bR\bfm{\varPsi}(-x_1,x_2,-x_3).
\eeq
Substituting these expressions into the Betti formula \eq{biq_mech+}, and replacing $\bU(x_1,x_2, x_3)$ with $\bU(x_1^{'}-x_1, x_2, x_3^{'}-x_3)$ which corresponds 
to a shift within the plane $(x_1,x_3)$, for the upper half-plane we obtain: 
\begin{displaymath}
    \int_{(x_2=0^+)}  \left[\bR\bU(x_{1}^{'}-x_1,0^+,x_{3}^{'}-x_3)\cdot\BGs_2(x_1,0^+,x_3)\right.
 \end{displaymath}
 \begin{displaymath}
\left.-\bR \bfm{\varSigma}_2(x_{1}^{'}-x_1,0^+,x_{3}^{'}-x_3)\cdot\bu(x_1,0^+,x_3)\right]\ dx_1dx_3 
\end{displaymath}
\begin{displaymath}
  +  \gamma_t^+ \int_{(x_2=0^+)}\frac{\partial \varPhi(x_{1}^{'}-x_1,0^+,x_{3}^{'}-x_3)}{\partial x_2}\theta(x_1,0^+,x_3)\ dx_1dx_3 
 \end{displaymath}
\begin{displaymath}
 +  \beta_t^+ \int_{(x_2=0^+)}\varPhi(x_{1}^{'}-x_1,0^+,x_{3}^{'}-x_3)q_2^{(1)}(x_1,0^+,x_3) \ dx_1dx_3 
\end{displaymath}
\begin{displaymath}
   +  \gamma_c^+ \int_{(x_2=0^+)}\frac{\partial \varPhi(x_{1}^{'}-x_1,0^+,x_{3}^{'}-x_3)}{\partial x_2}\chi(x_1,0^+,x_3)\ dx_1dx_3
    \end{displaymath}
\beq
 + \beta_c^+  \int_{(x_2=0^+)} \varPhi(x_{1}^{'}-x_1,0^+,x_{3}^{'}-x_3)j_2^{(1)}(x_1,0^+,x_3)\ dx_1dx_3=0.
 \label{IIplus}
 \eeq
An analogous expression is derived for the lower half-plane where $0^{+}$ is replaced with $0^{-}$. Subtracting this expression from the (\ref{IIplus}), we obtain
\begin{eqnarray}
 &   & \int_{(x_2=0)}  \left\{\bR\jump{0.1}{\bU}(x_{1}^{'}-x_1,x_{3}^{'}-x_3)\cdot\langle\BGs_2\rangle(x_1,x_3)\right.\nonumber\\
 & + & \left. \bR\langle\bU\rangle(x_{1}^{'}-x_1,x_{3}^{'}-x_3)\cdot\jump{0.1}{\BGs_2}(x_1,x_3)
 -\bR \langle\bfm{\varSigma}_2\rangle(x_{1}^{'}-x_1,x_{3}^{'}-x_3)\cdot\jump{0.1}{\bu}(x_1,x_3)\right\} \ dx_1dx_3  \nonumber \\
 & + & \int_{(x_2=0)}\left\{\left\langle\frac{\partial \varPhi_{\gamma t}}{\partial x_2}\right\rangle(x_{1}^{'}-x_1,x_{3}^{'}-x_3)\jump{0.1}{\theta}(x_1,x_3)
 +\bjump{0.35}{\frac{\partial \varPhi_{\gamma t}}{\partial x_2}}(x_{1}^{'}-x_1,x_{3}^{'}-x_3)\langle\theta\rangle(x_1,x_3)\right.\nonumber\\
 & + & \left. \jump{0.1}{\varPhi_{\beta t}}(x_{1}^{'}-x_1,x_{3}^{'}-x_3)\left\langle q_2 \right\rangle(x_1,x_3)
 + \langle\varPhi_{\beta t}\rangle(x_{1}^{'}-x_1,x_{3}^{'}-x_3)\jump{0.1}{q_2}(x_{1},x_3)\right\} \ dx_1\nonumber\\
 & + & \int_{(x_2=0)}\left\{\left\langle\frac{\partial \varPhi_{\gamma c}}{\partial x_2}\right\rangle(x_{1}^{'}-x_1,x_{3}^{'}-x_3)\jump{0.1}{\chi}(x_1,x_3)
 +\bjump{0.35}{\frac{\partial \varPhi_{\gamma c}}{\partial x_2}}(x_{1}^{'}-x_1,x_{3}^{'}-x_3)\langle\chi\rangle(x_1,x_3)\right.\nonumber\\
 & + & \left. \jump{0.1}{\varPhi_{\beta c}}(x_{1}^{'}-x_1,x_{3}^{'}-x_3)\left\langle j_2 \right\rangle(x_1,x_3)\
 + \langle\varPhi_{\beta c}\rangle(x_{1}^{'}-x_1,x_{3}^{'}-x_3)\jump{0.1}{j_2}(x_{1},x_3)\right\} \ dx_1dx_3=0,\nonumber\\
 \label{rec_id1}
 \end{eqnarray}
where the notation $\varPhi_{\gamma t}, \varPhi_{\gamma c}, \varPhi_{\beta t}$ and $\varPhi_{\beta c}$ indicates the following normalized potentials:
\beq
\varPhi_{\gamma t}=\gamma_t\varPhi, \quad \varPhi_{\gamma c}=\gamma_c\varPhi, \quad \varPhi_{\beta t}=\beta_t\varPhi, \quad \varPhi_{\beta c}=\beta_c\varPhi.
\label{Phi_norm}
\eeq
The physical quantitities $\bu, \BGs_2, \theta, q_2, \chi, j_2$ involved in the \eq{rec_id1} can be represented as follows
\beq
\label{repr}
f(x_1,x_3)=f^{(+)}(x_1,x_3)+f^{(-)}(x_1,x_3),
\eeq
where the superscripts $^{(+)}$ and $^{(-)}$ denote functions whose support is restricted to the positive and negative semi-axes, respectively:
\beq
f^{(+)}(x_1,x_3)=f(x_1,x_3)H(x_1), \quad f^{(-)}(x_1,x_3)=f(x_1,x_3)H(-x_1),
\eeq
and $H$ denotes the Heaviside function. The reciprocity identity \eq{rec_id1} then becomes
\begin{eqnarray}
  &   & \bR\jump{0.1}{\bU}\circledast\langle\BGs_2\rangle^{(+)}-\bR \langle\bfm{\varSigma}_2\rangle\circledast\jump{0.1}{\bu}^{(-)} \nonumber\\
= & - & \bR\jump{0.1}{\bU}\circledast\langle\BGs_2\rangle^{(-)}-\bR\langle\bU\rangle\circledast\jump{0.1}{\BGs_2}^{(-)}-\bR\langle\bU\rangle\circledast\jump{0.1}{\BGs_2}^{(+)}
+\bR \langle\bfm{\varSigma}_2\rangle\circledast\jump{0.1}{\bu}^{(+)} \nonumber\\
  & - & \left\langle\frac{\partial \varPhi_{\gamma t}}{\partial x_2}\right\rangle\circledast\jump{0.1}{\theta}^{(-)}-
  \bjump{0.35}{\frac{\partial \varPhi_{\gamma t}}{\partial x_2}}\circledast\langle\theta\rangle^{(-)}-
 \left\langle\frac{\partial \varPhi_{\gamma t}}{\partial x_2}\right\rangle\circledast\jump{0.1}{\theta}^{(+)}-
 \bjump{0.35}{\frac{\partial \varPhi_{\gamma t}}{\partial x_2}}\circledast\langle\theta\rangle^{(+)}\nonumber\\
  & - & \jump{0.1}{\varPhi_{\beta t}}\circledast\left\langle q_2 \right\rangle^{(-)}
  -\langle\varPhi_{\beta t}\rangle\circledast\jump{0.1}{q_2}^{(-)}-
  \jump{0.1}{\varPhi_{\beta t}}\circledast\left\langle q_2 \right\rangle^{(+)}
  -\langle \varPhi_{\beta t} \rangle\circledast\jump{0.1}{q_2}^{(+)}\nonumber\\
  & - & \left\langle\frac{\partial \varPhi_{\gamma c}}{\partial x_2}\right\rangle\circledast\jump{0.1}{\chi}^{(-)}
  -\bjump{0.35}{\frac{\partial \varPhi_{\gamma c}}{\partial x_2}}\circledast\langle\chi\rangle^{(-)}-
 \left\langle\frac{\partial \varPhi_{\gamma c}}{\partial x_2}\right\rangle\circledast\jump{0.1}{\chi}^{(+)}
 -\bjump{0.35}{\frac{\partial \varPhi_{\gamma c}}{\partial x_2}}\circledast\langle\chi\rangle^{(+)}\nonumber\\
  & - & \jump{0.1}{\varPhi_{\beta c}}\circledast\left\langle j_2 \right\rangle^{(-)}
  -\langle\varPhi_{\beta c}\rangle\circledast\jump{0.1}{j_2}^{(-)}
  -\jump{0.1}{\varPhi_{\beta c}}\circledast\left\langle j_2 \right\rangle^{(+)}
  -\langle\varPhi_{\beta c}\rangle\circledast\jump{0.1}{j_2}^{(+)},
\label{Betti2}
  \end{eqnarray}
  where the symbol $\circledast$ denotes the convolution with respect to both $x_1$ and $x_3$, which is defined as follows \citep{Arfk1}:
  \beq
  f \circledast g=\iint_{-\infty}^{+\infty}f(x_{1}^{'}-x_1,x_{3}^{'}-x_3)g(x_1, x_3)dx_1dx_3=0,
  \eeq
  
On the basis of representation \eq{repr}, $\langle\BGs_2\rangle^{(+)}$ is the traction along the interface, ahead of the crack tip, and $\jump{0.1}{\bu}^{(-)}$ is the crack opening 
(displacement discontinuity across the crack faces). $\jump{0.1}{\bU}$ and $\langle \bU \rangle$ are the symmetrical and skew-
symmetrical weight functions matrices defined and derived in closed form in \citet{PiccMish1}, whereas the term $\langle \bfm{\varSigma}_2\rangle$ 
stands for the traction along the $x_1-$axis  corresponding to the singular auxiliary displacements $\bU$.

The integral equation \eq{Betti2} is the generalization to thermoelastic diffusive media of the Betti identity derived in \citet{PiccMish3} and \citet{MorPicc2}, and it relates the physical
solution $\bu, \BGs_2, \theta, \chi$ to the auxiliary singular solution $\bU, \bfm{\varSigma}_2$. Note that \eq{Betti2} is valid for the most general case of static interfacial crack problems 
in presence of thermal and diffusive effects, and includes also the case of imperfect mechanical interface conditions, associated with the discontinuity of tractions and  displacements
at the interface ahead of the crack tip \citep{VelMish1}. In the sequel, perfect contact conditions at the interface for $x_1>0$ will be assumed for mechanical fields,
then: $\jump{0.1}{\bu}^{(+)}=\jump{0.1}{\BGs_2}^{(+)}=0$, whereas according to definitions reported in Section \ref{thermohalf}
discontinuity of the thermodiffusive quantities along the whole plane $x_2=0$ will be considered (imperfect thermodiffusive interface conditions). Moreover, 
the notation \eq{load} is used to indicate the symmetrical and skew-symmetrical mechanical loads applied at the crack faces, then 
$\langle\BGs_2\rangle^{(-)}=\langle\bp\rangle$ and $\jump{0.1}{\BGs_2}^{(-)}=\jump{0.1}{\bp}$, respectively. 
The integral identity \eq{Betti2} becomes:
\begin{eqnarray}
  &   & \bR\jump{0.1}{\bU}\circledast\langle\BGs_2\rangle^{(+)}-\bR \langle\bfm{\varSigma}_2\rangle\circledast\jump{0.1}{\bu}^{(-)}
  = \ -\bR\jump{0.1}{\bU}\circledast\langle\bp\rangle-\bR\langle\bU\rangle\circledast\jump{0.1}{\bp} \nonumber\\
  & - & \left\langle\frac{\partial \varPhi_{\gamma t}}{\partial x_2}\right\rangle\circledast\jump{0.1}{\theta}-\bjump{0.35}{\frac{\partial \varPhi_{\gamma t}}{\partial x_2}}\circledast\langle\theta\rangle
  -\langle\varPhi_{\beta t}\rangle\circledast\jump{0.1}{ q_1}-\jump{0.1}{\varPhi_{\beta t}}\circledast\left\langle q_2 \right\rangle\nonumber\\
  & - & \left\langle\frac{\partial \varPhi_{\gamma c}}{\partial x_2}\right\rangle\circledast\jump{0.1}{\chi}-\bjump{0.35}{\frac{\partial \varPhi_{\gamma c}}{\partial x_2}}\circledast\langle\chi\rangle
  -\langle\varPhi_{\beta c}\rangle\circledast\jump{0.1}{j_2}-\jump{0.1}{\varPhi_{\beta c}}\circledast\left\langle j_2 \right\rangle,
\label{Betti3}
  \end{eqnarray}
 where the terms $\langle\theta\rangle, \langle\chi\rangle, \langle q_2 \rangle $ and $\langle j_2 \rangle $ stand respectively for the mean values of the temperature, mass concentration,
 heat flux and particle current on the plane $x_2=0$, defined respectively by expressions \eq{average_td} and \eq{average_flux}. Similarly,
 $\jump{0.1}{\theta}, \jump{0.1}{\chi}, \jump{0.1}{p_2}$ and $\jump{0.1}{j_2}$ denotes the jumps of these functions across the same plane $x_2=0$, and defined by \eq{jump_td} and \eq{jump_flux}.
 
 The Betti's identity \eq{Betti3} will be used further in the text in order to formulate the illustrated crack problem at the interface between 
 dissimilar thermodiffusive materials in terms of singular integral equations. Explicit integral identities relating the loading applied at the crack faces, the temperature,
 the mass concentration, the heat and mass fluxes to the resulting crack opening and traction ahead of the crack tip are derived for the two-dimensional case. 
\section{Integral identities}
\label{ident}
In this Section, starting from the general identity \eq{Betti3}, the problem of a two-dimensional interfacial crack in presence of thermodiffusion
is reduced to a system of explicit integral equations. The solution of these equations provides exact expressions for the crack opening and for the tractions ahead 
of the tip corresponding to an arbitrary loading configuration and to the temperature and concentration profiles of the system, obtained by solving the Poisson's equations \eq{poisson}. 
Since thermal conduction and diffusion are both assumed to be isotropic, only normal stresses are induced by temperature changes and mass diffusion 
(see constitutive relation \eq{hooke}). For this reason, in the considered two-dimensional problem antiplane stress and deformations are not affected by thermodiffusion, so that
only the plane strain case is studied in order to estimate the contribution of thermal and diffusive stresses on interface fracture phenomena.

For plane strain deformations, the Betti identity \eq{Betti3} relating the physical solution $\bu=[u_1, u_2]^T$ and $\BGs_2=[\Gs_{21}, \Gs_{22}]^T$ with 
the weight function $\bU$ and $\bfm{\varSigma}_2$ becomes
\begin{eqnarray}
   &   & \bR\jump{0.1}{\bU}\ast\langle\BGs_2\rangle^{(+)}-\bR \langle\bfm{\varSigma}_2\rangle\ast\jump{0.1}{\bu}^{(-)}
= \ -\bR\jump{0.1}{\bU}\ast\langle\bp\rangle-\bR\langle\bU\rangle\ast\jump{0.1}{\bp} \nonumber\\
  & - & \left\langle\frac{\partial \varPhi_{\gamma t}}{\partial x_2}\right\rangle\ast\jump{0.1}{\theta}-\bjump{0.35}{\frac{\partial \varPhi_{\gamma t}}{\partial x_2}}\ast\langle\theta\rangle
  -\langle\varPhi_{\beta t}\rangle\ast\jump{0.1}{ q_1}-\jump{0.1}{\varPhi_{\beta t}}\ast\left\langle q_2 \right\rangle\nonumber\\
  & - & \left\langle\frac{\partial \varPhi_{\gamma c}}{\partial x_2}\right\rangle\ast\jump{0.1}{\chi}-\bjump{0.35}{\frac{\partial \varPhi_{\gamma c}}{\partial x_2}}\ast\langle\chi\rangle
  -\langle\varPhi_{\beta c}\rangle\ast\jump{0.1}{j_2}-\jump{0.1}{\varPhi_{\beta c}}\ast\left\langle j_2 \right\rangle,
\label{Betti_plane}
  \end{eqnarray}
where $\langle\bp\rangle=[\langle p_1\rangle, \langle p_2\rangle]^T$, $\jump{0.1}{\bp}=[\jump{0.1}{p_1}, \jump{0.1}{p_2}]^T$, and the symbol $\ast$ denotes the convolution
with respect to the variable $x_1$. Here and in the sequel of the article we will used the following $2\times 2$ matrices:
\beq
\bR = 
\left[
\begin{array}{cc}
-1 & 0 \\  0 & 1 
\end{array}
\right], \quad
\bI = 
\left[
\begin{array}{cc}
1 & 0 \\  0 & 1 
\end{array}
\right],  \quad
\bE = 
\left[
\begin{array}{cc}
0 & 1 \\  -1 & 0
\end{array}
\right], \quad
\bF = 
\left[
\begin{array}{cc}
0 & 1 \\  1 & 0 
\end{array}
\right].
\eeq

For plane elastic bimaterials, two linearly independent weight functions, $\bU^j=[U_1^j,U_2^j]^T,\bfm{\varSigma}_2=[\varSigma_1^j,\varSigma_2^j]^T, j=1,2$ are needed in order to define a complete
basis of the space of singular solutions of the homogeneous problem (see \cite{PiccMish1} and \cite{MorRad1} for details). The weight functions tensors may be constructed by ordering
the components of each independent solution in columns:
\beq
\bU = 
\left[
\begin{array}{cc}
U_{1}^{1} & U_{1}^{2} \\ U_{2}^{1}  & U_{2}^{2} 
\end{array}
\right], \quad
\bfm{\varSigma}_2=
\left[
\begin{array}{cc}
\varSigma_{1}^{1} & \varSigma_{1}^{2} \\ \varSigma_{2}^{1}  & \varSigma_{2}^{2} 
\end{array}
\right].
\label{Umatrices}
\eeq

In order to express the weight function matrix \eq{Umatrices}$_{(1)}$ in terms of elastic Lam\'e potentials \eq{PhiPsi}, it is important to note that 
in the considered plane strain case the vector potential $\bfm{\varPsi}$ possesses only one non-zero component directed along $x_3-$axis, and then:
$\bfm{\varPsi}=\varPsi_3\bfm{e}_3=\varPsi\bfm{e}_3$. Two couples of elastic potentials  $(\varPhi^1,\varPsi^1)$ and $(\varPhi^2,\varPsi^2)$ are introduced, 
and the components of $\bU$ are expressed as follows:
\beq
U_{1}^{1}=\frac{\partial \varPhi^1}{\partial x_1}+\frac{\partial \varPsi^1}{\partial x_2}, \quad U_{2}^{1}=\frac{\partial \varPhi^1}{\partial x_2}-\frac{\partial \varPsi^1}{\partial x_1}, \quad
U_{1}^{2}=\frac{\partial \varPhi^2}{\partial x_1}+\frac{\partial \varPsi^2}{\partial x_2}, \quad U_{2}^{2}=\frac{\partial \varPhi^2}{\partial x_2}-\frac{\partial \varPsi^2}{\partial x_1}.
\label{U2pot}
\eeq

Let us introduce the Fourier transform of a generic function $f$ with respect to the variable $x_1$:
\beq
\tilde{f}(\xi) = \mF[f(x_1)] = \int_{-\infty}^\infty f(x_1) e^{i \xi x_{1}} dx_{1}, \quad 
f(x_1) = \mF^{-1}[\tilde{f}(\xi)] = \frac{1}{2\pi} \int_{-\infty}^\infty \tilde{f}(\xi) e^{-i \xi x_{1}} d\xi.
\label{fourier}
\eeq
Applying \eq{fourier} to expression \eq{Betti_plane} and considering representations \eq{U2pot}, we get 
\begin{eqnarray}
  &   & \jump{0.1}{\tilde{\bU}}^{T}\bR\langle\tilde{\BGs}_2\rangle^{+}-\langle\tilde{\bfm{\varSigma}}_2\rangle^{T}\bR\jump{0.1}{\tilde{\bu}}^{-} 
= \ -\jump{0.1}{\tilde{\bU}}^{T}\bR\langle\tilde{\bp}\rangle-\langle\tilde{\bU}\rangle^{T}\bR\jump{0.1}{\tilde{\bp}} \nonumber\\
  & - & \left\langle\frac{\partial \tilde{\bfm{\varPhi}}_{\gamma t}}{\partial x_2}\right\rangle\jump{0.1}{\tilde{\theta}}
  -\bjump{0.35}{\frac{\partial \tilde{\bfm{\varPhi}}_{\gamma t}}{\partial x_2}}\langle\tilde{\theta}\rangle
  -\langle\tilde{\bfm{\varPhi}}_{\beta t}\rangle\jump{0.1}{\tilde{q}_2}
  -\jump{0.1}{\tilde{\bfm{\varPhi}}_{\beta t}}\left\langle \tilde{q}_2 \right\rangle\nonumber\\
  & - & \left\langle\frac{\partial \tilde{\bfm{\varPhi}}_{\gamma c}}{\partial x_2}\right\rangle\jump{0.1}{\tilde{\chi}}
  -\bjump{0.35}{\frac{\partial \tilde{\bfm{\varPhi}}_{\gamma c}}{\partial x_2}}\langle\tilde{\chi}\rangle
  -\langle\tilde{\bfm{\varPhi}}_{\beta c}\rangle\jump{0.1}{\tilde{j}_2}
  -\jump{0.1}{\tilde{\bfm{\varPhi}}_{\beta c}}\left\langle \tilde{j}_2 \right\rangle,
\label{Betti_fourier}
  \end{eqnarray}
where $\bfm{\varPhi}_{\gamma t}=[\gamma_t \varPhi^1, \gamma_t \varPhi^2]^T, \bfm{\varPhi}_{\beta t}=[\beta_t \varPhi^1, \beta_t \varPhi^2]^T, \bfm{\varPhi}_{\gamma c}=[\gamma_c \varPhi^1,
\gamma_c \varPhi^2]^T, \bfm{\varPhi}_{\beta c}=[\beta_c \varPhi^1, \beta_c \varPhi^2]^T$, whereas the superscripts $^+$ and $^-$ denote $``+"$ and $``-"$ functions respectively. 
In order to obtain from expression \eq{Betti_fourier} a system of integral equations relating 
the crack opening and the traction ahead of the tip with applied loading and thermodiffusive quantities, explicit expressions for the elastic potentials $\varPhi^1$ and $\varPhi^2$ are needed.

\subsection{Derivation of elastic potentials}
Applying the Fourier transform to equations \eq{U2pot}, the following system of non-homogeneous ODEs is derived:
\beq
\left\{
\begin{array}{c}
\tilde{\varPsi}^{'}-i\xi\tilde{\varPhi}=\tilde{U}_{1}  \\  
\tilde{\varPhi}^{'}+i\xi\tilde{\varPsi}=\tilde{U}_{2},
\end{array}
\right .
\label{ODE}
\eeq
where ${'}$ denotes the total derivative with respect to $x_2$. Since both the pairs of transformed elastic potentials $(\tilde{\varPhi}^1, \tilde{\varPsi}^1)$ and $(\tilde{\varPhi}^2, 
\tilde{\varPsi}^2)$ can be obtained by solving the system \eq{ODE} considering respectively $\tilde{\bU}^1=[\tilde{U}^1_1, \tilde{U}^1_2]^T$ and $\tilde{\bU}^2=
[\tilde{U}^2_1, \tilde{U}^2_2]^T$ as non-homogeneous parts, the superscripts $^1$ and $^2$ has been omitted. 

The Fourier transform of the singular displacements $\tilde{U}_1$ and $\tilde{U}_2$ was derived in \citet{PiccMish1}. Considering the upper half-plane $x_1>0$, they are given by
\begin{eqnarray}
 \tilde{U}_1(\xi,x_2) & = & \left\{\left[x_2-\frac{\lambda^{+} + 2\mu^+}{|\xi|(\lambda^{+}+\mu^+)}\right]\tilde{\varSigma}_{21}^{-}+
 i\left[\frac{\mu^+}{\xi(\lambda^{+}+\mu^{+})}-\sign(\xi)x_2 \right]\tilde{\varSigma}_{22}^{-}\right\}\frac{e^{-|\xi|x_2}}{2\mu^+},\label{U1}\\
 \tilde{U}_2(\xi,x_2) & = & \left\{-i\left[\sign(\xi)x_2+\frac{\mu^+}{\xi(\lambda^{+}+\mu^{+})}\right]\tilde{\varSigma}_{21}^{-}-
 \left[x_2+\frac{\lambda^{+}+2\mu^{+}}{|\xi|(\lambda^{+}+\mu^{+})}\right]\tilde{\varSigma}_{22}^{-}\right\}\frac{e^{-|\xi|x_2}}{2\mu^+},\label{U2}
\end{eqnarray}
where $\tilde{\varSigma}_{21}^{-}(\xi)=\tilde{\varSigma}_{21}^{-}(\xi,0^-)=\tilde{\varSigma}_{21}^{-}(\xi,0^+)$ and 
$\tilde{\varSigma}_{22}^{-}(\xi)=\tilde{\varSigma}_{22}^{-}(\xi,0^-)=\tilde{\varSigma}_{22}^{-}(\xi,0^+)$ are Fourier transform of the traction at the interface (see for example \citet{MorRad1}).
For the lower half-plane, replacing $-|\xi|$ with $|\xi|$, $\mu^+$ with $\mu^-$ and $\lambda^+$ with $\lambda^-$, similar expressions are found. The singular
displacements \eq{U1} and \eq{U2} are substituted into \eq{ODE}, and the solution of the linear system yields the following expressions for the transformed elastic potentials, valid
for the upper-half plane (see Appendix B for details):
\begin{eqnarray}
 \tilde{\varPhi}(\xi,x_2) & = & \left\{-i\frac{x_2\mu^+}{\xi(\lambda^{+}+\mu^+)}\tilde{\varSigma}_{21}^{-}+
 \left[\frac{1}{2\xi^2}-\frac{x_2\mu^+}{|\xi|(\lambda^{+}+\mu^{+})}\right]\tilde{\varSigma}_{22}^{-}\right\}\frac{e^{-|\xi|x_2}}{2\mu^+},\nonumber\\
 \tilde{\varPsi}(\xi,x_2) & = & \left\{-\frac{x_2(\lambda^{+}+2\mu^{+})}{|\xi|(\lambda^{+}+\mu^{+})}\tilde{\varSigma}_{21}^{-}
 +i\left[\frac{\sign(\xi)}{2\xi^2}+\frac{x_2(\lambda^{+}+2\mu^{+})}{\xi(\lambda^{+}+\mu^{+})}\right]\tilde{\varSigma}_{22}^{-}\right\}\frac{e^{-|\xi|x_2}}{2\mu^+}.\nonumber\\
\label{PhiPsiexpl}
 \end{eqnarray}
Similarly to the case of the displacements, replacing $-|\xi|$ with $|\xi|$, $\mu^+$ with $\mu^-$ and $\lambda^+$ with $\lambda^-$ in the \eq{PhiPsiexpl} the elastic potentials for the 
lower half-plane are obtained. 

It is possible now to derive the jump and the average of the Fourier transforms of the normalized elastic potentials \eq{Phi_norm} across the plane containing the crack. The 
traces of the expressions \eq{PhiPsiexpl} on the plane $x_2=0$ containing the crack are given by 
\beq
\tilde{\varPhi}(\xi,0^+)=
\left[
\begin{array}{cc}
0,
\ \cfrac{1}{4\xi^2\mu^+}
\end{array}
\right] 
\left[
\begin{array}{c}
 \tilde{\varSigma}_{21}^{-} \\
 \tilde{\varSigma}_{22}^{-}
\end{array}
\right], \quad
\tilde{\varPsi}(\xi,0^+)=
\left[
\begin{array}{cc}
0,
\ \cfrac{i\sign(\xi)}{4\xi^2\mu^+}
\end{array}
\right] 
\left[
\begin{array}{c}
 \tilde{\varSigma}_{21}^{-} \\
 \tilde{\varSigma}_{22}^{-}
\end{array}
\right],
\eeq

\beq
\tilde{\varPhi}(\xi,0^-)=
\left[
\begin{array}{cc}
0 ,
\ \cfrac{1}{4\xi^2\mu^-}
\end{array}
\right] 
\left[
\begin{array}{c}
 \tilde{\varSigma}_{21}^{-} \\
 \tilde{\varSigma}_{22}^{-}
\end{array}
\right], \quad
\tilde{\varPsi}(\xi,0^-)=
\left[
\begin{array}{cc}
0,
\ -\cfrac{i\sign(\xi)}{4\xi^2\mu^-}
\end{array}
\right] 
\left[
\begin{array}{c}
 \tilde{\varSigma}_{21}^{-} \\
 \tilde{\varSigma}_{22}^{-}
\end{array}
\right],
\eeq
 so that, taking into account the two linearly independent weight functions defined by means of equations \eq{Umatrices}, 
 we derive the following matrix form for the traces of $\varPhi^1$ and $\varPhi^2$:
\beq
\left[
\begin{array}{c}
\tilde{\varPhi}^1(\xi,0^+)\\
\tilde{\varPhi}^2(\xi,0^+)
\end{array}
\right]=
\left[
\begin{array}{cc}
 \tilde{\varSigma}_{21}^{1-} &  \tilde{\varSigma}_{22}^{1-}\\
 \tilde{\varSigma}_{21}^{2-} &  \tilde{\varSigma}_{22}^{2-}
\end{array}
\right]
\left[
\begin{array}{c}
0 \\
\cfrac{1}{4\xi^2\mu^+}
\end{array}
\right]=\langle\tilde{\bfm{\varSigma}}_2\rangle^{T}\bfm{\eta}^+(\xi),
\label{Phi0+}
\eeq
\beq
\left[
\begin{array}{c}
\tilde{\varPhi}^1(\xi,0^-)\\
\tilde{\varPhi}^2(\xi,0^-)
\end{array}
\right]=
\left[
\begin{array}{cc}
 \tilde{\varSigma}_{21}^{1-} &  \tilde{\varSigma}_{22}^{1-}\\
 \tilde{\varSigma}_{21}^{2-} &  \tilde{\varSigma}_{22}^{2-}
\end{array}
\right]
\left[
\begin{array}{c}
0 \\
\cfrac{1}{4\xi^2\mu^-}
\end{array}
\right]=\langle\tilde{\bfm{\varSigma}}_2\rangle^{T}\bfm{\eta}^-(\xi).
\label{Phi0-}
\eeq
Using equations \eq{Phi0+} and \eq{Phi0-} together with the definition \eq{Phi_norm} the jump and the average of the Fourier transform of the normalized potentials $\bfm{\varPhi}_{\beta t}$ become:
\beq
\jump{0.1}{\tilde{\bfm{\varPhi}}_{\beta t}}=\langle\tilde{\bfm{\varSigma}}_2\rangle^{T}(\beta^+_t\bfm{\eta}^+-\beta^-_t\bfm{\eta}^-)=
\langle\tilde{\bfm{\varSigma}}_2\rangle^{T}\jump{0.1}{\bfm{\eta}_t},
\label{jumpT}
\eeq
\beq
\langle\tilde{\bfm{\varPhi}}_{\beta t}\rangle=\frac{1}{2}\langle\tilde{\bfm{\varSigma}}_2\rangle^{T}(\beta^+_t\bfm{\eta}^++\beta^-_t\bfm{\eta}^-)=
\langle\tilde{\bfm{\varSigma}}_2\rangle^{T}\langle\bfm{\eta}_t \rangle,
\eeq
\beq
\jump{0.1}{\tilde{\bfm{\varPhi}}_{\beta c}}=\langle\tilde{\bfm{\varSigma}}_2\rangle^{T}(\beta^+_c\bfm{\eta}^+-\beta^-_c\bfm{\eta}^-)=
\langle\tilde{\bfm{\varSigma}}_2\rangle^{T}\jump{0.1}{\bfm{\eta}_c},
\eeq
\beq
\langle\tilde{\bfm{\varPhi}}_{\beta c}\rangle=\frac{1}{2}\langle\tilde{\bfm{\varSigma}}_2\rangle^{T}(\beta^+_c\bfm{\eta}^++\beta^-_c\bfm{\eta}^-)=
\langle\tilde{\bfm{\varSigma}}_2\rangle^{T}\langle\bfm{\eta}_c \rangle,
\label{aveC}
\eeq
where: 
\beq
\jump{0.1}{\bfm{\eta}_t}= 
\frac{1}{\xi^2}\left[
\begin{array}{c}
 0 \\
 h_t
\end{array}
\right],\
\langle \bfm{\eta}_t \rangle= 
\frac{1}{2\xi^2}\left[
\begin{array}{c}
 0 \\
 \ell_t
\end{array}
\right], \
\jump{0.1}{\bfm{\eta}_c}= 
\frac{1}{\xi^2}\left[
\begin{array}{c}
 0 \\
 h_c
\end{array}
\right],\
\langle \bfm{\eta}_c \rangle= 
\frac{1}{2\xi^2}\left[
\begin{array}{c}
 0 \\
 \ell_c
\end{array}
\right],
\label{eta_c}
\eeq
and the expressions for the bimaterial thermodiffusive parameters $h_t, \ell_t, h_c$ and $\ell_c$ are reported in Appendix A. 

In order to derive explicit integral identities by equation \eq{Betti_fourier}, also expressions for the jump and the average of the Fourier transform of the heat and mass fluxes
across of the plane containing the crack are required. The derivative respect to $x_2$ of the transformed elastic potentials in the upper half-plane \eq{PhiPsiexpl} is given by 
\begin{displaymath}
 \tilde{\varPhi}^{'}(\xi,x_2) = \left\{i\left[\frac{x_2\mu^{+}\sign(\xi)}{\lambda^{+}+\mu^{+}}-\frac{\mu^{+}}{\xi(\lambda^{+}+\mu^{+})}\right]\tilde{\varSigma}_{21}^{-}+
 \left[\frac{x_2\mu^{+}}{\lambda^{+}+\mu^{+}}-\frac{\lambda^{+}+
 3\mu^{+}}{2|\xi|(\lambda^{+}+\mu^{+})}\right]\tilde{\varSigma}_{22}^{-}\right\}\frac{e^{-|\xi|x_2}}{2\mu^+},
 \label{PhiPsiexpl'}
 \end{displaymath}
 \begin{displaymath}
 \tilde{\varPsi}^{'}(\xi,x_2)=\left\{\left[\frac{x_2(\lambda^{+}+2\mu^{+})}{\lambda^{+}+\mu^{+}}-\frac{\lambda^{+}+
 2\mu^{+}}{|\xi|(\lambda^{+}+\mu^{+})}\right]\tilde{\varSigma}_{21}^{-}-i\left[\frac{x_2(\lambda^{+}+2\mu^{+})\sign(\xi)}{\lambda^{+}+\mu^{+}}-\frac{\lambda^{+}+
 3\mu^{+}}{2\xi(\lambda^{+}+\mu^{+})}\right]\tilde{\varSigma}_{22}^{-}\right\}\frac{e^{-|\xi|x_2}}{2\mu^+},\nonumber\\
\end{displaymath}
replacing $-|\xi|$ with $|\xi|$, $\mu^+$ with $\mu^-$ and $\lambda^+$ with $\lambda^-$ in these expressions, the derivatives with respect to $x_2$ of the trasformed elastic potentials for the 
lower half-plane are obtained. The traces of $\tilde{\varPhi}^{'}$ and $\tilde{\varPsi}^{'}$ on the plane $x_2=0$ then become
\beq
\tilde{\varPhi}^{'}(\xi,0^+)=
\left[
\begin{array}{cc}
-\cfrac{i}{2(\lambda^{+}+\mu^{+})\xi}, &
\ -\cfrac{\lambda^{+}+3\mu^{+}}{4\mu^{+}(\lambda^{+}+\mu^{+})|\xi|}
\end{array}
\right] 
\left[
\begin{array}{c}
 \tilde{\varSigma}_{21}^{-} \\
 \tilde{\varSigma}_{22}^{-}
\end{array}
\right],
\eeq
\beq
\tilde{\varPsi}^{'}(\xi,0^+)=
\left[
\begin{array}{cc}
-\cfrac{\lambda^{+}+2\mu^{+}}{2\mu^{+}(\lambda^{+}+\mu^{+})|\xi|}, &
\ \cfrac{i(\lambda^{+}+3\mu^{+})}{4\mu^{+}(\lambda^{+}+\mu^{+})\xi}
\end{array}
\right] 
\left[
\begin{array}{c}
 \tilde{\varSigma}_{21}^{-} \\
 \tilde{\varSigma}_{22}^{-}
\end{array}
\right],
\eeq
\beq
\tilde{\varPhi}^{'}(\xi,0^-)=
\left[
\begin{array}{cc}
-\cfrac{i}{2(\lambda^{-}+\mu^{-})\xi}, &
\ \cfrac{\lambda^{-}+3\mu^{-}}{4\mu^{-}(\lambda^{-}+\mu^{-})|\xi|}
\end{array}
\right] 
\left[
\begin{array}{c}
 \tilde{\varSigma}_{21}^{-} \\
 \tilde{\varSigma}_{22}^{-}
\end{array}
\right],
\eeq
\beq
\tilde{\varPsi}^{'}(\xi,0^-)=
\left[
\begin{array}{cc}
\cfrac{\lambda^{-}+2\mu^{-}}{2\mu^{-}(\lambda^{-}+\mu^{-})|\xi|}, &
\ \cfrac{i(\lambda^{-}+3\mu^{-})}{4\mu^{-}(\lambda^{-}+\mu^{-})\xi}
\end{array}
\right] 
\left[
\begin{array}{c}
 \tilde{\varSigma}_{21}^{-} \\
 \tilde{\varSigma}_{22}^{-}
\end{array}
\right],
\eeq
and the following matrix representation can be introduced:
\beq
\left[
\begin{array}{c}
\tilde{\varPhi}^{1'}(\xi,0^+)\\
\tilde{\varPhi}^{2'}(\xi,0^+)
\end{array}
\right]=
\left[
\begin{array}{cc}
 \tilde{\varSigma}_{21}^{1-} &  \tilde{\varSigma}_{22}^{1-}\\
 \tilde{\varSigma}_{21}^{2-} &  \tilde{\varSigma}_{22}^{2-}
\end{array}
\right]
\left[
\begin{array}{c}
-\cfrac{i}{2(\lambda^{+}+\mu^{+})\xi}\\
-\cfrac{\lambda^{+}+3\mu^{+}}{4\mu^{+}(\lambda^{+}+\mu^{+})|\xi|}
\end{array}
\right]=\langle\tilde{\bfm{\varSigma}}_2\rangle^{T}\bfm{\zeta}^+(\xi),
\eeq
\beq
\left[
\begin{array}{c}
\tilde{\varPhi}^{1'}(\xi,0^-)\\
\tilde{\varPhi}^{2'}(\xi,0^-)
\end{array}
\right]=
\left[
\begin{array}{cc}
 \tilde{\varSigma}_{21}^{1-} &  \tilde{\varSigma}_{22}^{1-}\\
 \tilde{\varSigma}_{21}^{2-} &  \tilde{\varSigma}_{22}^{2-}
\end{array}
\right]
\left[
\begin{array}{c}
-\cfrac{i}{2(\lambda^{-}+\mu^{-})\xi}\\
\cfrac{\lambda^{-}+3\mu^{-}}{4\mu^{-}(\lambda^{-}+\mu^{-})|\xi|}
\end{array}
\right]=\langle\tilde{\bfm{\varSigma}}_2\rangle^{T}\bfm{\zeta}^-(\xi).
\eeq

The jump and the average of the Fourier transform of the derivatives of the elastic potentials across the plane $x_2$ can be finally written as 
\beq
\bjump{0.35}{\frac{\partial \tilde{\bfm{\varPhi}}_{\gamma t}}{\partial x_2}}=\langle\tilde{\bfm{\varSigma}}_2\rangle^{T}(\gamma^+_t\bfm{\zeta}^+-\gamma^-_t\bfm{\zeta}^-)=
\langle\tilde{\bfm{\varSigma}}_2\rangle^{T}\jump{0.1}{\bfm{\zeta}_t},
\label{jumpFt}
\eeq
\beq
\left\langle\frac{\partial \tilde{\bfm{\varPhi}}_{\gamma t}}{\partial x_2}\right\rangle=\frac{1}{2}\langle\tilde{\bfm{\varSigma}}_2\rangle^{T}(\gamma^+_t\bfm{\zeta}^++
\gamma^-_t\bfm{\zeta}^-)=\langle\tilde{\bfm{\varSigma}}_2\rangle^{T}\langle\bfm{\zeta}_t \rangle,
\eeq
\beq
\bjump{0.35}{\frac{\partial \tilde{\bfm{\varPhi}}_{\gamma c}}{\partial x_2}}=\langle\tilde{\bfm{\varSigma}}_2\rangle^{T}(\gamma^+_c\bfm{\zeta}^+-\gamma^-_c\bfm{\zeta}^-)=
\langle\tilde{\bfm{\varSigma}}_2\rangle^{T}\jump{0.1}{\bfm{\zeta}_c},
\eeq
\beq
\left\langle\frac{\partial \tilde{\bfm{\varPhi}}_{\gamma c}}{\partial x_2}\right\rangle=\frac{1}{2}\langle\tilde{\bfm{\varSigma}}_2\rangle^{T}(\gamma^+_c\bfm{\zeta}^++
\gamma^-_c\bfm{\zeta}^-)=\langle\tilde{\bfm{\varSigma}}_2\rangle^{T}\langle\bfm{\zeta}_c \rangle,
\label{aveFc}
\eeq
where:
\beq
\jump{0.1}{\bfm{\zeta}_t}= 
-\frac{1}{\xi}\left[
\begin{array}{c}
 im_t \\
 n_t\sign(\xi)
\end{array}
\right],\
\langle \bfm{\zeta}_t \rangle= 
-\frac{1}{2\xi}\left[
\begin{array}{c}
 ip_t \\
 q_t\sign(\xi)
\end{array}
\right],
\label{zeta_t}
\eeq
\beq
\jump{0.1}{\bfm{\zeta}_c}= 
-\frac{1}{\xi}\left[
\begin{array}{c}
 im_c \\
 n_c\sign(\xi)
\end{array}
\right],\
\langle \bfm{\zeta}_c \rangle= 
-\frac{1}{2\xi}\left[
\begin{array}{c}
 ip_c \\
 q_c\sign(\xi)
\end{array}
\right],
\label{zeta_c}
\eeq
and the bimaterial thermodiffusive parameters $m_t, n_t, p_t, q_t, m_c, n_c, p_c$ and $q_c$ are defined in Appendix A. 

The obtained expressions for the jump and the average of the Fourier transform of elastic potentials and fluxes across of the plane $x_2=0$ are now used together with the Betti identity 
\eq{Betti_fourier} for reducing the interface crack problem to a system of explicit singular integral equations. 
\subsection{Explicit integral identities}
Substituting expressions \eq{jumpT}-\eq{aveC} and \eq{jumpFt}-\eq{aveFc} in equation \eq{Betti_fourier}, we obtain:
\begin{eqnarray}
  &   & \jump{0.1}{\tilde{\bU}}^{T}\bR\langle\tilde{\BGs}_2\rangle^{+}-\langle\tilde{\bfm{\varSigma}}_2\rangle^{T}\bR\jump{0.1}{\tilde{\bu}}^{-} 
= \ -\jump{0.1}{\tilde{\bU}}^{T}\bR\langle\tilde{\bp}\rangle-\langle\tilde{\bU}\rangle^{T}\bR\jump{0.1}{\tilde{\bp}} \nonumber\\
  & - & \langle\tilde{\bfm{\varSigma}}_2\rangle^{T}\langle\bfm{\zeta}_t \rangle\jump{0.1}{\tilde{\theta}}
  -\langle\tilde{\bfm{\varSigma}}_2\rangle^{T}\jump{0.1}{\bfm{\zeta}_t}\langle\tilde{\theta}\rangle
  -\langle\tilde{\bfm{\varSigma}}_2\rangle^{T}\langle\bfm{\eta}_t \rangle\jump{0.1}{\tilde{q}_2}
  -\langle\tilde{\bfm{\varSigma}}_2\rangle^{T}\jump{0.1}{\bfm{\eta}_t}\left\langle\tilde{q}_2\right\rangle\nonumber\\
  & - & \langle\tilde{\bfm{\varSigma}}_2\rangle^{T}\langle\bfm{\zeta}_c \rangle\jump{0.1}{\tilde{\chi}}
  -\langle\tilde{\bfm{\varSigma}}_2\rangle^{T}\jump{0.1}{\bfm{\zeta}_c}\langle\tilde{\chi}\rangle
  -\langle\tilde{\bfm{\varSigma}}_2\rangle^{T}\langle\bfm{\eta}_c \rangle\jump{0.1}{\tilde{j}_2}
  -\langle\tilde{\bfm{\varSigma}}_2\rangle^{T}\jump{0.1}{\bfm{\eta}_c}\left\langle\tilde{j}_2\right\rangle.
\label{Betti_mod}
  \end{eqnarray}
Multiplying both sides by $\bR^{-1}\jump{0.1}{\tilde{\bU}}^{-T}$, we get
\begin{eqnarray}
  &   & \langle \tilde{\BGs}_2\rangle^{+}-\bB\jump{0.1}{\tilde{\bu}}^{-} = -\langle\tilde{\bp}\rangle-\bA\jump{0.1}{\tilde{\bp}}
  - \langle\bg_t \rangle\jump{0.1}{\tilde{\theta}}-\jump{0.1}{\bg_t}\langle\tilde{\theta}\rangle
  -\langle\bd_t \rangle\jump{0.1}{\tilde{q}_2}
  -\jump{0.1}{\bd_t}\left\langle\tilde{q}_2\right\rangle\nonumber\\
  &  & -\langle\bg_c \rangle\jump{0.1}{\tilde{\chi}}-\jump{0.1}{\bg_c}\langle\tilde{\chi}\rangle
  -\langle\bd_c \rangle\jump{0.1}{\tilde{j}_2}
  -\jump{0.1}{\bd_c}\left\langle\tilde{j}_2\right\rangle,
  \label{Betti_mod2}
\end{eqnarray}
where $\bA$ and $\bB$ are the following matrices, identical to those introduced in \citet{PiccMish3} and \citet{MorPicc2} for the elastic problem without thermodiffusion:
\beq
\bA=\bR^{-1}\jump{0.1}{\tilde{\bU}}^{-T}\langle\tilde{\bU}\rangle^{T}\bR, \quad \bB=\bR^{-1}\jump{0.1}{\tilde{\bU}}^{-T}\langle\tilde{\bfm{\varSigma}}_2\rangle^{T}\bR,
\label{AB}
\eeq
the vectors $\langle\bg_t \rangle, \jump{0.1}{\bg_t}, \langle\bd_t \rangle$ and $\jump{0.1}{\bd_t}$, associate to thermal effects, are given:
\beq
\langle\bg_t \rangle= \bM\langle\bfm{\zeta}_t \rangle, \ \ \jump{0.1}{\bg_t}=\bM\jump{0.1}{\bfm{\zeta}_t}, \ \ \langle\bd_t \rangle=\bM\langle\bfm{\eta}_t \rangle, \ \ 
\jump{0.1}{\bd_t}= \bM\jump{0.1}{\bfm{\eta}_t},
\label{t_quant}
\eeq
and finally the terms $\langle\bg_c \rangle, \jump{0.1}{\bg_c}, \langle\bd_c \rangle$ and $\jump{0.1}{\bd_c}$, corresponding to the contribution of the mass diffusion, are defined as follows:
\beq
\langle\bg_c \rangle= \bM\langle\bfm{\zeta}_c \rangle, \ \ \jump{0.1}{\bg_c}=\bM\jump{0.1}{\bfm{\zeta}_c}, \ \ \langle\bd_c \rangle=\bM\langle\bfm{\eta}_c \rangle, \ \
\jump{0.1}{\bd_c}= \bM\jump{0.1}{\bfm{\eta}_c},
\label{e_quant}
\eeq
where the matrix $\bM$ is given by
\beq
\label{M}
\bM=\bR^{-1}\jump{0.1}{\tilde{\bU}}^{-T}\langle\tilde{\bfm{\varSigma}}_2\rangle^{T}.
\eeq

The matrices $\bA$, $\bB$ and $\bM$ can be easily computed using the symmetric and skew-symmetric weight functions derived by \citet{PiccMish1}:
\beq
\jump{0.1}{\tilde{\bU}}=-\frac{1}{|\xi|}[b\bI-id\sign(\xi)\bE]\langle\tilde{\bfm{\varSigma}}_2\rangle, \quad 
\langle\tilde{\bU}\rangle=-\frac{b}{2|\xi|}[\alpha\bI-i\gamma\sign(\xi)\bE]\langle\tilde{\bfm{\varSigma}}_2\rangle,
\label{weight_expl}
\eeq
where $\alpha$ and $d/b$ are the so-called Dundurs parameters, $b$ and $\gamma$ are bimaterial constants defined in Appendix A.
Using expressions \eq{weight_expl} together with the definitions \eq{AB} and \eq{M}, we get
\beq
 \bA  =  \frac{b}{2(b^2-d^2)}[(\alpha b -\gamma d)\bI+i(\alpha d-\gamma b)\sign(\xi)\bE],
 \eeq
 \beq
 \bB  =  -\frac{|\xi|}{b^2-d^2}[b \bI+id\sign(\xi)\bE],\qquad
 \bM  =  -\frac{|\xi|}{b^2-d^2}[b \bR+id \sign(\xi)\bF]. \label{Mexpl}
\eeq

Substituting \eq{eta_c} and \eq{M} in equation \eq{t_quant}, the explicit expressions for the 
vectors $\jump{0.1}{\bd_t},\langle \bd_t \rangle, \jump{0.1}{\bg_t}$ and $\langle \bg_t \rangle$ are obtained:
\beq
\jump{0.1}{\bd_t}= 
-\frac{h_t}{(b^2-d^2)|\xi|}\left[
\begin{array}{c}
 id\sign(\xi) \\
 b
\end{array}
\right],\
\langle \bd_t \rangle= 
-\frac{\ell_t}{2(b^2-d^2)|\xi|}\left[
\begin{array}{c}
 id\sign(\xi) \\
 b
\end{array}
\right],
\eeq
\beq
\jump{0.1}{\bg_t}= 
\frac{1}{b^2-d^2}\left[
\begin{array}{c}
 i(dn_t-bm_t)\sign(\xi) \\
 bn_t-dm_t
\end{array}
\right],\
\langle \bg_t \rangle= 
\frac{1}{2(b^2-d^2)}\left[
\begin{array}{c}
 i(dq_t-bq_t)\sign(\xi) \\
 bq_t-dp_t
\end{array}
\right].
\eeq
Similarly, for $\jump{0.1}{\bd_c},\langle \bd_c \rangle, \jump{0.1}{\bg_c}$ and $\langle \bg_c \rangle$ we get
\beq
\jump{0.1}{\bd_c}= 
-\frac{h_c}{(b^2-d^2)|\xi|}\left[
\begin{array}{c}
 id\sign(\xi) \\
 b
\end{array}
\right],\
\langle \bd_c \rangle= 
-\frac{\ell_t}{2(b^2-d^2)|\xi|}\left[
\begin{array}{c}
 id\sign(\xi) \\
 b
\end{array}
\right],
\eeq
\beq
\jump{0.1}{\bg_c}= 
\frac{1}{b^2-d^2}\left[
\begin{array}{c}
 i(dn_c - bm_c)\sign(\xi) \\
 bn_c-dm_c
\end{array}
\right],\
\langle \bg_c \rangle= 
\frac{1}{2(b^2-d^2)}\left[
\begin{array}{c}
 i(dq_c-bp_c)\sign(\xi) \\
 bq_c-dp_c
\end{array}
\right].
\eeq

Since explicit expressions for all the terms of the identity \eq{Betti_mod2} have been derived, we can now apply the inverse Fourier transform to this equation, obtaining two distinct 
relationships corresponding to the two cases $x_1<0$ and $x_1>0$:
\begin{eqnarray}
  &   & \langle\bp\rangle +\mF_{x_1<0}^{-1}[\bA\jump{0.1}{\tilde{\bp}}]+ \mF_{x_1<0}^{-1}[\langle\bg_t \rangle\jump{0.1}{\tilde{\theta}}] 
  +\mF_{x_1<0}^{-1}[\jump{0.1}{\bg_t}\langle\tilde{\theta}\rangle]\nonumber\\
  &  &+\mF_{x_1<0}^{-1}\left[\langle\bd_t \rangle\jump{0.1}{ \tilde{q}_2}\right] 
  +\mF_{x_1<0}^{-1}\left[\jump{0.1}{\bd_t}\left\langle\tilde{q}_2\right\rangle\right]
  +\mF_{x_1<0}^{-1}[\langle\bg_c \rangle\jump{0.1}{\tilde{\chi}}]+\mF_{x_1<0}^{-1}[\jump{0.1}{\bg_c}\langle\tilde{\chi}\rangle]\nonumber\\
  & & +\mF_{x_1<0}^{-1}\left[\langle\bd_c \rangle\jump{0.1}{\tilde{j}_2}\right]
  +\mF_{x_1<0}^{-1}\left[\jump{0.1}{\bd_c}\left\langle \tilde{j}_2\right\rangle\right]=\mF_{x_1<0}^{-1}[\bB\jump{0.1}{\tilde{\bu}}^-],
\label{Betti_inv1}
  \end{eqnarray}
\begin{eqnarray}
  &   &  \langle \BGs_2 \rangle^{(+)}+\mF_{x_1>0}^{-1}[\bA\jump{0.1}{\tilde{\bp}}]+ \mF_{x_1>0}^{-1}[\langle\bg_t \rangle\jump{0.1}{\tilde{\theta}}] 
  +\mF_{x_1>0}^{-1}[\jump{0.1}{\bg_t}\langle\tilde{\theta}\rangle]\nonumber\\
  &  &+\mF_{x_1>0}^{-1}\left[\langle\bd_t \rangle\jump{0.1}{\tilde{q}_2}\right] 
  +\mF_{x_1>0}^{-1}\left[\jump{0.1}{\bd_t}\left\langle \tilde{q}_2\right\rangle\right]
  +\mF_{x_1>0}^{-1}[\langle\bg_c \rangle\jump{0.1}{\tilde{\chi}}^{-}]+\mF_{x_1>0}^{-1}[\jump{0.1}{\bg_c}\langle\tilde{\chi}\rangle]\nonumber\\
  & & +\mF_{x_1>0}^{-1}\left[\langle\bd_c \rangle\jump{0.1}{\tilde{j}_2}\right]
  +\mF_{x_1>0}^{-1}\left[\jump{0.1}{\bd_c}\left\langle\tilde{j}_2\right\rangle\right]=\mF_{x_1>0}^{-1}[\bB\jump{0.1}{\tilde{\bu}}^-].
\label{Betti_inv2}
  \end{eqnarray}
Note that the term $\langle \tilde{\BGs_2} \rangle^{+}$ in the identity \eq{Betti_mod2} cancels from the \eq{Betti_inv1} because it is a $``+"$ function, while $\langle\tilde{\bp}\rangle$
cancels from the \eq{Betti_inv2} because it is a $``-"$ function. To proceed further, the following inverse Fourier transforms are evaluated by means 
of the convolutions theorem and distributions theory \citep{Roos1, BrychPrud}:
\beq
\mF^{-1}[|\xi|\tilde{f}(\xi)]=\frac{1}{\pi x_1}\ast\frac{\partial f}{\partial x_1}, \quad \mF^{-1}[\sign(\xi)\tilde{f}(\xi)]=-\frac{i}{\pi x_1}
\ast f(x_1),
\label{inv_first}
\eeq
\beq
\mF^{-1}\left[\frac{1}{\xi}\tilde{f}(\xi)\right]=-\frac{i}{2}\sign{x_1}\ast f(x_1), 
\quad \mF^{-1}\left[\frac{1}{|\xi|}\tilde{f}(\xi)\right]=-\frac{1}{\pi}(\gamma_{eul}+\ln|x_1|)\ast
f(x_1), 
\label{inv_last}
\eeq
where the function $f$ can be both the jump and the average of $\theta, \chi, q_2$ and $j_2$, and $\gamma_{eul}$ is the Euler's gamma constant. Using the inverse transforms \eq{inv_first}-\eq{inv_last} in equations \eq{Betti_inv1} and \eq{Betti_inv2} we finally obtain the explicit 
integral identities for plane cracks problems at the interface between dissimilar thermodiffusive elastic materials:
\begin{eqnarray}
  &  & \langle\bp\rangle +\bmA^{(s)}\jump{0.1}{\bp}+ \bmG^{(s-)}_{1t}\jump{0.1}{\theta} 
  +\bmG^{(s-)}_{2t}\langle\theta\rangle+\bmD^{(s-)}_{1t}\jump{0.1}{q_2} 
  +\bmD^{(s-)}_{2t}\left\langle q_2\right\rangle\nonumber\\
  &  &+\bmG^{(s-)}_{1c}\jump{0.1}{\chi}+\bmG^{(s-)}_{2c}\langle\chi\rangle+\bmD^{(s-)}_{1c}\jump{0.1}{j_2}
  +\bmD^{(s-)}_{2c}\left\langle j_2\right\rangle = \ \bmB^{(s)}\frac{\partial \jump{0.1}{\bu}^{(-)}}{\partial x_1}, \qquad  x_1<0,
\label{id1}
  \end{eqnarray}
\begin{eqnarray}
  &  &  \langle \BGs_2 \rangle^{(+)} +\bmA^{(c)}\jump{0.1}{\bp}+ \bmG^{(s+)}_{1t}\jump{0.1}{\theta}
  +\bmG^{(s+)}_{2t}\langle\theta\rangle+\bmD^{(s+)}_{1t}\jump{0.1}{q_2}
  +\bmD^{(s+)}_{2t}\left\langle q_2 \right\rangle\nonumber\\
  &  &+\bmG^{(c)}_{1c}\jump{0.1}{\chi}^{(-)}+\bmG^{(s+)}_{2c}\langle\chi\rangle+\bmD^{(s+)}_{1c}\jump{0.1}{j_2}
  +\bmD^{(s+)}_{2c}\left\langle j_2 \right\rangle = \ \bmB^{(c)}\frac{\partial \jump{0.1}{\bu}^{(-)}}{\partial x_1}, \qquad  x_1>0.
\label{id2}
  \end{eqnarray}
$\bmA^{(s)}, \bmB^{(s)}: F(\fR_{-})\rightarrow F(\fR_{-})$ and $\bmA^{(c)}, \bmB^{(c)}: F(\fR_{-})\rightarrow F(\fR_{+})$ are respectively singular and compact matrix operators 
\citep{Gakh1,GohKr1,Krein1}, where $F(\fR_{\pm})$ is some functional space of functions defined on $\fR_{\pm}$. These operators are given in the form:
\beq
\bmA^{(s)}=\frac{b}{2(b^2-d^2)}[(b\alpha-d\gamma)\bI+(d\alpha-b\gamma)\bE\mS^{(s)}], \quad \bmB^{(s)}=-\frac{1}{b^2-d^2}[b\bI\mS^{(s)}-d\bE],
\label{opAs}
\eeq
\beq
\bmA^{(c)}=\frac{b(d\alpha-b\gamma)}{2(b^2-d^2)}\bE\mS^{(c)}, \quad \bmB^{(c)}=-\frac{b}{b^2-d^2}\bI\mS^{(c)}.
\label{opABc}
\eeq
The singular operator $\mS^{(s)}=\mP_{-}\mS\mP_{-}$ and the compact operator $\mS^{(c)}=\mP_{+}\mS\mP_{-}$ are defined in details in \citet{PiccMish3} and \citet{MorPicc2} by introducing the
singular operator of the Cauchy type $\mS$:
\beq
\mS f=\frac{1}{\pi x_1}\ast f, 
\eeq
and the orthogonal projectors $\mP_{\pm}$:
\beq
\mP_{\pm}f=\left\{
\begin{array}{c}
f(x_1) \quad \pm x_1 \geq 0, \\  
0 \qquad \ \ \ \ \mbox{otherwise}.
\end{array}
\right .
\eeq

The vector operators $\bmG^{(s+),(s-)}_{1t}, \bmG^{(s+),(s-)}_{2t}, \bmG^{(s+),(s-)}_{1c}$ and $\bmG^{(s+),(s-)}_{2c}$, associate to the contributions of the temperature and of the concentration
on the plane containing the crack $x_2=0$, depend by $\mS^{(s-)}$ and $\mS^{(s+)}$:
\beq
\bmG^{(s-)}_{1t}= 
\frac{1}{2(b^2-d^2)}\left[
\begin{array}{c}
 (dq_t-bp_t)\mS^{(s-)} \\
 bq_t-dp_t
\end{array}
\right],\
\bmG^{(s-)}_{2t}= 
\frac{1}{b^2-d^2}\left[
\begin{array}{c}
 (dn_t - bm_t)\mS^{(s-)} \\
 bn_t -dm_t
\end{array}
\right],
\label{op_TD1}
\eeq
\beq
\bmG^{(s-)}_{1c}= 
\frac{1}{2(b^2-d^2)}\left[
\begin{array}{c}
 (dq_c - bp_c)\mS^{(s-)} \\
 bq_c -dp_c
\end{array}
\right],\
\bmG^{(s-)}_{2c}= 
\frac{1}{b^2-d^2}\left[
\begin{array}{c}
 (dn_c - bm_c)\mS^{(s-)} \\
 bn_c -dm_c
\end{array}
\right],
\eeq
\beq
\bmG^{(s+)}_{1t}= 
\frac{1}{2(b^2-d^2)}\left[
\begin{array}{c}
 (dq_t - bp_t)\mS^{(s+)} \\
 bq_t-dp_t
\end{array}
\right],\
\bmG^{(s+)}_{2t}= 
\frac{1}{b^2-d^2}\left[
\begin{array}{c}
 (dn_t - bm_t)\mS^{(s+)} \\
 bn_t-dm_t
\end{array}
\right],
\eeq
\beq
\bmG^{(s+)}_{1c}= 
\frac{1}{2(b^2-d^2)}\left[
\begin{array}{c}
 (dq_c - bp_c)\mS^{(s+)} \\
 bq_c -dp_c
\end{array}
\right],\
\bmG^{(s+)}_{2c}= 
\frac{1}{b^2-d^2}\left[
\begin{array}{c}
 (dm_c - bn_c)\mS^{(s+)} \\
 bn_c-dm_c
\end{array}
\right],
\label{op_TD2}
\eeq
where $\mS^{(s-)}=\mP_{-}\mS$ and $\mS^{(s+)}=\mP_{+}\mS$ are singular integral operators. $\bmD^{(s+),(s-)}_{1t}, \bmD^{(s+),(s-)}_{2t}, \bmD^{(s+),(s-)}_{1c}$ and $\bmD^{(s+),(s-)}_{2c}$, 
related to the heat and mass fluxes at the interface are given by 
\beq
\bmD^{(s-)}_{1t}= 
-\frac{\ell_t}{4(b^2-d^2)}\left[
\begin{array}{c}
 d\mK^{(s-)} \\
 b\mJ^{(s-)}
\end{array}
\right],\
\bmD^{(s-)}_{2t}= 
-\frac{h_t}{2(b^2-d^2)}\left[
\begin{array}{c}
 d\mK^{(s-)} \\
 b\mJ^{(s-)}
\end{array}
\right],
\label{op_FLUX1}
\eeq
\beq
\bmD^{(s-)}_{1c}= 
-\frac{\ell_c}{4(b^2-d^2)}\left[
\begin{array}{c}
 d\mK^{(s-)} \\
 b\mJ^{(s-)}
\end{array}
\right],\
\bmD^{(s-)}_{2c}= 
-\frac{h_c}{2(b^2-d^2)}\left[
\begin{array}{c}
 d\mK^{(s-)} \\
 b\mJ^{(s-)}
\end{array}
\right],
\eeq
\beq
\bmD^{(s+)}_{1t}= 
-\frac{\ell_t}{4(b^2-d^2)}\left[
\begin{array}{c}
 d\mK^{(s+)} \\
 b\mJ^{(s+)}
\end{array}
\right],\
\bmD^{(s+)}_{2t}= 
-\frac{h_t}{2(b^2-d^2)}\left[
\begin{array}{c}
 d\mK^{(s+)} \\
 b\mJ^{(s+)}
\end{array}
\right],
\label{op_FLUX1+}
\eeq
\beq
\bmD^{(s+)}_{1c}= 
-\frac{\ell_c}{4(b^2-d^2)}\left[
\begin{array}{c}
 d\mK^{(s+)} \\
 b\mJ^{(s+)}
\end{array}
\right],\
\bmD^{(s+)}_{2c}= 
-\frac{h_c}{2(b^2-d^2)}\left[
\begin{array}{c}
 d\mK^{(s+)} \\
 b\mJ^{(s+)}
\end{array}
\right],
\label{op_FLUX2}
\eeq
where $\mK$ and $\mJ$ are the integral operators
\beq
\mK f =\sign(x_1)\ast f, \quad  \mJ f=-\frac{2}{\pi}(\gamma_{eul}+\ln|x_1|)\ast f,
\eeq
and $\mK^{(s-)}=\mP_{-}\mK,\ \mK^{(s+)}=\mP_{+}\mK, \ \mJ^{(s-)}=\mP_{-}\mJ,\ \mJ^{(s+)}=\mP_{+}\mJ$.

Equations \eq{id1} and \eq{id2} form a system of integral equations relating the crack opening and the traction ahead of the tip to the mechanical loading applied at the crack faces and to 
the values of temperature, mass concentration, heat and mass fluxes on the plane containing the fracture. The solution of \eq{id1} by the inversion of the matrix operator $\bmB^{(s)}$
provides crack opening that corresponds to arbitrary loading configurations acting on the faces and arbitrary profiles of the thermodiffusive quantities.
The obtained expressions for $\jump{0.1}{\bu}^{(-)}$ can then be used in \eq{id2} for evaluating the traction ahead of the crack tip. 

Note that, in the case where the Dundurs parameter $d$ vanishes, the operators \eq{opAs}-\eq{opABc} can be written as
\beq
\bmA^{(s)}=\frac{\alpha}{2}\bI-\frac{\gamma}{2}\bE\mS^{(s)}, \quad \bmB^{(s)}=-\frac{1}{b}\bI\mS^{(s)}, \quad
\bmA^{(c)}=\frac{\gamma}{2}\bE\mS^{(c)}, \quad \bmB^{(c)}=-\frac{1}{b}\bI\mS^{(c)},
\label{opdec2}
\eeq
and then the integral equations \eq{id1} can be decoupled and reduced to:
\begin{eqnarray}
  &  & -\frac{1}{b}\mS^{(s)}\frac{\partial \jump{0.1}{\bu}^{(-)}}{\partial x_1}=\langle\bp\rangle +\left(\frac{\alpha}{2}\bI-\frac{\gamma}{2}\bE\mS^{(s)}\right)\jump{0.1}{\bp}\
  + \bmG^{(s-)}_{1t}\jump{0.1}{\theta} 
  +\bmG^{(s-)}_{2t}\langle\theta\rangle+\bmD^{(s-)}_{1t}\jump{0.1}{q_2} \nonumber\\
  & &+\bmD^{(s-)}_{2t}\left\langle q_2\right\rangle
  +\bmG^{(s-)}_{1c}\jump{0.1}{\chi}+\bmG^{(s-)}_{2c}\langle\chi\rangle+\bmD^{(s-)}_{1c}\jump{0.1}{j_2}
  +\bmD^{(s-)}_{2c}\left\langle j_2\right\rangle, \qquad  x_1<0.
\label{id1dec}
\end{eqnarray}
In the case of homogeneous material, we additionaly have $\alpha=0$ for the operators \eq{opdec2}, $q_t=m_t=q_c=m_c=0$ for the \eq{op_TD1}-\eq{op_TD2} and $h_t=h_c=0$ for the
\eq{op_FLUX1}-\eq{op_FLUX2}, and the integral equations \eq{id1dec} become
\beq
\left\{
\begin{array}{ll}
  -\cfrac{1}{b}\mS^{(s)}\cfrac{\partial \jump{0.1}{u_1}^{(-)}}{\partial x_1}=\langle p_1\rangle-\cfrac{\gamma}{2}\mS^{(s)}\jump{0.1}{p_2}-\cfrac{p_t}{2b}\mS^{(s-)}\jump{0.1}{\theta}
  -\cfrac{p_c}{2b}\mS^{(s-)}\jump{0.1}{\chi}, &           \\
                                              &     x_1<0.  \\
  -\cfrac{1}{b}\mS^{(s)}\cfrac{\partial \jump{0.1}{u_2}^{(-)}}{\partial x_1}=\langle p_2\rangle+\cfrac{\gamma}{2}\mS^{(s)}\jump{0.1}{p_1}+\cfrac{n_t}{b}\langle\theta\rangle
  +\cfrac{n_c}{b}\langle\chi\rangle-\cfrac{\ell_t}{4b}\mJ^{(s-)}\jump{0.1}{q_2}-\cfrac{\ell_c}{4b}\mJ^{(s-)}\jump{0.1}{j_2} , &
\end{array}
\right.
\label{id1hom}
\eeq
Assuming symmetrical mechanical load ($\jump{0.1}{p_1}=\jump{0.1}{p_2}=0$) together with negligible thermodiffusive effects, equations \eq{id1hom} become identical to expressions 
derived by \citet{Rice1}.

Differently from the general case \eq{id1}, where the inversion of the matrix operator $\bmB^{(s)}$ by means of the general procedure reported in \citet{Vekua1} is required, 
for the solution of equations \eq{id1dec} and \eq{id1hom} only the inversion of the operator $\mS^{(s)}$ (see \citet{PiccMish3}) is needed. In order to allow some simple examples of
application of the obtained integral identities, in the next Section uncoupled integral identities 
\eq{id1dec} will be solved for certain illustrative cases, and effects of temperature and heat fluxes profile on crack opening and traction ahead of the tip will be discussed in details.

\section{Illustrative examples}
\label{examples}
In this section, the derived general integral identities are applied in order to study the effects of two particular temperature and heat flux profiles
on crack opening and traction ahead of the tip. The temperature and heat flux functions used in the illustrative examples here proposed are derived
in Appendix D by the solution of the Laplace's equation in a semi-plane considering two different sets of boundary conditions. Note that,
due to the same form of the associate vector integral operators (see expressions \eq{op_TD1}-\eq{op_TD2} and \eq{op_FLUX1}-\eq{op_FLUX2}), 
the results here reported for the temperature field can be immediately generalized for studying the effects of mass concentration and mass flux variations
at the interface.
\subsection{Punctually localized heat flux profile at the interface}
\label{exlog}
\begin{figure}[!htcb]
\centering
\includegraphics[width=13.6cm]{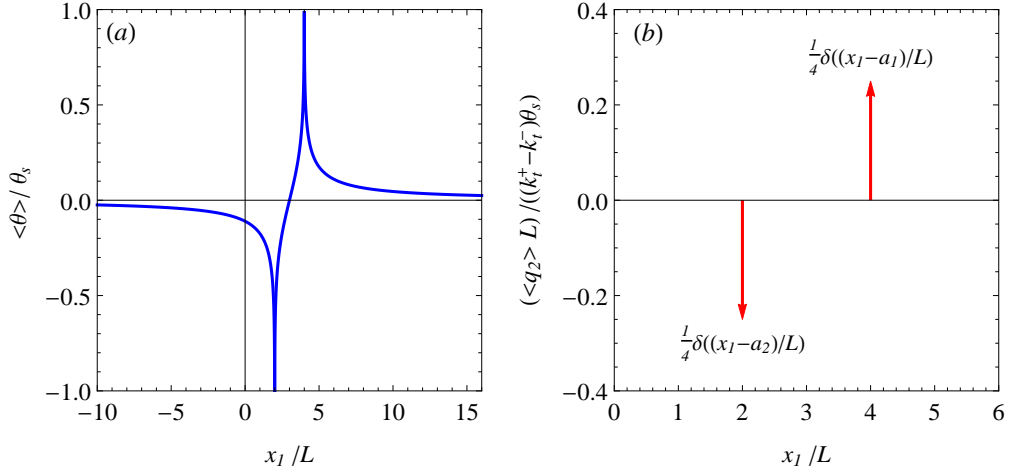}
\caption{\footnotesize (a): Normalized temperature profile at the interface $x_2=0$ reported for $a_{1}/L=4$ and  $a_{2}/L=2$;
(b): Normalized average heat flux profile at the interfaces $x_2=0$ reported for $a_{1}/L=4$ and  $a_{2}/L=2$.}
\label{figthetaqlog}
\end{figure}

The following punctually localized profiles for the avegare and the jump of normal heat flux across the plane $x_2=0$, containing both the crack and the interface, are assumed:
\beq
\left\langle q_1 \right\rangle (x_1)=\frac{\theta_s}{4 L}(k_t^+ -k_t^-)\left[\delta\left(\frac{x_1-a_1}{L}\right)-\delta\left(\frac{x_1-a_2}{L}\right)\right],
\label{qlog1}
\eeq
\beq
 \jump{0.1}{q_1}(x_1)=\frac{\theta_s}{2 L}(k_t^+ +k_t^-)\left[\delta\left(\frac{x_1+a_1}{L}\right)-\delta\left(\frac{x_1-a_2}{L}\right)\right],
\label{qlog2}
 \eeq
 where $a_1, a_2, L>0$ and $a_1>a_2$. It can be easily verified that the functions \eq{qlog1} and \eq{qlog2} satisfy the integral balance conditions introduced in Section \ref{intercrack}. 
 The average and the jump of the temperature at $x_2=0$ corresponding to the heat flux profiles \eq{qlog1} and \eq{qlog2} can be evaluated solving the steady state 
 heat conduction equation for both the upper and the lower thermodiffusive half-space. This procedure is reported in details in Appendix \ref{tlogapp}, and the resulting average and jump of the
 temperature are given by
\beq
\left\langle \theta \right\rangle (x_1)=-\frac{\theta_s}{4\pi}\left[\ln(x_1-a_1)^2-\ln(x_1-a_2)^2\right],\quad \jump{0.1}{\theta}(x_1)=0.
\label{tlog}
\eeq
In Fig. \ref{figthetaqlog}, the variation of the normalized average temperature \eq{tlog} and of the normalized average heat flux \eq{qlog1} is plotted as a function 
of the spatial coordinate $x_1/L$ assuming $a_{1}/L=4$ and  $a_{2}/L=2$. 

For simplicity, the Dundurs parameter $d$ involved in the weight functions \eq{weight_expl} is assumed to be zero, and then the integral identities \eq{id1} and \eq{id2} decouple and equations
\eq{id1} take the form \eq{id1dec}. Considering only the contributions due to the fluxes and temperature profiles, the uncoupled integral identities \eq{id1dec} become
\begin{equation}
\left\{
\begin{array}{rrll}
 \mS^{(s)}\cfrac{\partial \jump{0.1}{u_1}^{(-)}}{\partial x_1} & = &  m_t\mS^{(s-)}\langle\theta\rangle,  &        \\
                                                               &   &                                     &  x_1<0\\
 \mS^{(s)}\cfrac{\partial \jump{0.1}{u_2}^{(-)}}{\partial x_1} & = & -n_t\langle\theta\rangle+\cfrac{h_t}{2}\mJ^{(s-)}\langle q_2 \rangle+\cfrac{l_t}{4}\mJ^{(s-)}\jump{0.1}{ q_2}.
\end{array}
\right.
\label{id-log}
\end{equation}
Remembering the definitions of $\mS^{(s-)}$ and $\mJ^{(s-)}$ introduced in the previous Section, and applying these operators to the heat and temperature profiles \eq{qlog1}, \eq{qlog2} and
\eq{tlog}, the following results are obtained:
\begin{equation}
 \mS^{(s-)}\langle\theta\rangle=-\frac{\theta_s}{4\pi}\mS^{(s-)}\left\{\ln\left[(x_1-a_1)^2\right] -\ln\left[(x_1-a_2)^2\right]\right\}=0,
\label{s-log}
 \end{equation}
\begin{align}
 \frac{h_t}{2}\mJ^{(s-)}\langle q_2 \rangle+\frac{l_t}{4}\mJ^{(s-)}\jump{0.1}{ q_2}& =\frac{\theta_s}{8L}\varUpsilon_t\mJ^{(s-)}
 \left[\delta\left(\frac{x_1-a_1}{L}\right)-\delta\left(\frac{x_1-a_2}{L}\right)\right]\nonumber\\
 & =-\frac{\theta_s}{8\pi}\varUpsilon_t\left\{\ln\left[(x_1-a_1)^2\right] -\ln\left[(x_1-a_2)^2\right]\right\}, 
\label{j-log}
 \end{align}
where $\varUpsilon_t=l_t (k_t^+ + k_t^-)+h_t (k_t^+ - k_t^-)$, and the explicit derivation of expression \eq{s-log} is reported in Appendix D. 

Substituting the \eq{s-log} and \eq{j-log} into equations \eq{id-log} we get
\begin{equation}
\left\{
\begin{array}{rrll}
 \mS^{(s)}\cfrac{\partial \jump{0.1}{u_1}^{(-)}}{\partial x_1} & = &  0,  &        \\
                                                               &   &                                     &  x_1<0\\
 \mS^{(s)}\cfrac{\partial \jump{0.1}{u_2}^{(-)}}{\partial x_1} & = & \cfrac{\theta_s}{4\pi}\bigg(n_t-\cfrac{\varUpsilon_t}{2}\bigg)\left\{\ln\left[(x_1-a_1)^2\right] -\ln\left[(x_1-a_2)^2\right]\right\}.
\end{array}
\right.
\end{equation}
The singular integral operator $\mS^{(s)}$ is now inverted following the procedure illustrated in \citet{PiccMish3} and \citet{MorPicc2}. This leads to
 \begin{align}
 \frac{\partial \jump{0.1}{u_1}^{(-)}}{\partial x_1} & = 0,               \\   
 \frac{\partial \jump{0.1}{u_2}^{(-)}}{\partial x_1} & = \frac{\theta_s}{\pi}\bigg(n_t-\cfrac{\varUpsilon_t}{2}\bigg)\left[-\sqrt{-\frac{a_1}{x_1}}+\sqrt{-\frac{a_2}{x_1}}+\arctan\sqrt{-\frac{a_1}{x_1}}-\arctan\sqrt{-\frac{a_2}{x_1}}\right],
 \end{align}
 which after integration gives explicit expressions for the crack opening associate to the heat fluxes \eq{qlog1} and \eq{qlog2} and to the temperature profiles at the interface \eq{tlog}:
  \begin{align}
  \jump{0.1}{u_1}^{(-)}(x_1) & = 0,               \\   
  \jump{0.1}{u_2}^{(-)}(x_1) & =  \frac{\theta_s}{\pi}\bigg(n_t-\cfrac{\Upsilon_t}{2}\bigg)\Bigg[\frac{\pi}{2}(a_1-a_2)+(\sqrt{-a_1x_1}-\sqrt{-a_2x_1}) \nonumber\\
                             &  \left.+(x_1-a_1)\arctan\sqrt{-\frac{a_1}{x_1}}-(x_1-a_2)\arctan\sqrt{-\frac{a_2}{x_1}}\right].
                             \label{ulog}
 \end{align}
 
 The tractions ahead of the crack tip are given by the solution of equations \eq{id2}, which in this case assume the form:
 \begin{equation}
\left\{
\begin{array}{rrll}
 \langle \Gs_{21} \rangle^{(+)} & = & -\cfrac{1}{b}\mS^{(c)}\cfrac{\partial \jump{0.1}{u_1}^{(-)}}{\partial x_1}+\cfrac{m_t}{b}\mS^{(s+)}\langle\theta\rangle,  &        \\
                                                               &   &                                     &  x_1>0\\
 \langle \Gs_{22} \rangle^{(+)} & = & -\cfrac{1}{b}\mS^{(c)}\cfrac{\partial \jump{0.1}{u_2}^{(-)}}{\partial x_1}-\cfrac{n_t}{b}
 \langle\theta\rangle+\cfrac{h_t}{2b}\mJ^{s+)}\langle q_2 \rangle+\cfrac{l_t}{4b}\mJ^{(s+)}\jump{0.1}{ q_2}.
\end{array}
\right.
\label{id2log}
\end{equation}
Applying the operators  $\mS^{(s+)}$ and $\mJ^{(s+)}$ to the flux and temperature functions  \eq{qlog1}, \eq{qlog2} and \eq{tlog}, the result is given by
\begin{equation}
 \mS^{(s+)}\langle\theta\rangle=-\frac{\theta_s}{4\pi}\mS^{(s+)}\left\{\ln\left[(x_1-a_1)^2\right] -\ln\left[(x_1-a_2)^2\right]\right\}=-\frac{\theta_s}{2}[H(a_1-x_1)-H(a_2-x_1)],
\label{s+log}
 \end{equation}
\begin{align}
 \frac{h_t}{2b}\mJ^{(s+)}\langle q_2 \rangle+\frac{l_t}{4b}\mJ^{(s+)}\jump{0.1}{ q_2} & =\frac{\theta_s}{8bL}\varUpsilon_t\mJ^{(s+)}
 \left[\delta\left(\frac{x_1-a_1}{L}\right)-\delta\left(\frac{x_1-a_2}{L}\right)\right]\nonumber\\
  & =-\frac{\theta_s}{8\pi b}\varUpsilon_t\left\{\ln\left[(x_1-a_1)^2\right] -\ln\left[(x_1-a_2)^2\right]\right\},
\label{j+log}
 \end{align}
 where $H(a_1-x_1)$ and $H(a_2-x_1)$ are Heaviside step functions (see Appendix D for details regarding the derivation of \eq{s+log}). Substituting expressions \eq{s+log} and \eq{j+log} into
 equations \eq{id2log}, the tractions ahead of the crack tip finally become
\begin{align}
 \langle \Gs_{21} \rangle^{(+)}(x_1) & = -\cfrac{m_t\theta_s}{2 b}[H(a_1-x_1)-H(a_2-x_1)], \label{sigmalog1}       \\
 \langle \Gs_{22} \rangle^{(+)}(x_1) & = -\cfrac{1}{b}\mS^{(c)}\cfrac{\partial \jump{0.1}{u_2}^{(-)}}{\partial x_1}+\cfrac{\theta_s}{4\pi b}\bigg(n_t-\cfrac{\varUpsilon_t}{2}\bigg)\left\{\ln\left[(x_1-a_1)^2\right] 
 -\ln\left[(x_1-a_2)^2\right]\right\}.
\label{sigmalog}
 \end{align}

 In order to study the influence of the different thermal expansion coefficients of the materials 
 on the crack opening and on the tractions ahead of the tip, the upper and lower half-space media are assumed to have identical values for the 
 elastic moduli $\lambda^+=\lambda^-=\lambda$, $\mu^+=\mu^-=\mu$, and for the thermal conductivity coefficients $k^+_t=k^-_t=k_t$, and dissimilar values for $\gamma_t$:
$\gamma_t^+\neq \gamma_t^-$. in this case, the bimaterial parameters involved in expressions \eq{ulog} and \eq{sigmalog} become:
\begin{displaymath}
m_t=\frac{\gamma_t^+}{2(\lambda+\mu)}\bigg(1-\frac{\gamma_t^-}{\gamma_t^+}\Big)=\varXi^+_t\bigg(1-\frac{\gamma_t^-}{\gamma_t^+}\bigg), \quad
 n_t-\frac{\varUpsilon_t}{2}=\frac{\gamma_t^+}{2(\lambda+\mu)}\bigg(1+\frac{\gamma_t^-}{\gamma_t^+}\Big)=\varXi^+_t\bigg(1+\frac{\gamma_t^-}{\gamma_t^+}\bigg),
\end{displaymath}
where $\varXi^+_t=\gamma_t^+/2(\lambda+\mu)$. Substituting these explicit constants in equations \eq{ulog} and \eq{sigmalog}, we obtain:
 \begin{align}
  \jump{0.1}{u_1}^{(-)}(x_1) & = 0,               \\   
  \jump{0.1}{u_2}^{(-)}(x_1) & =  \frac{\theta_s \varXi^+_t}{\pi}\bigg(1+\frac{\gamma_t^-}{\gamma_t^+}\bigg)\Bigg[\frac{\pi}{2}(a_1-a_2)+(\sqrt{-a_1x_1}-\sqrt{-a_2x_1}) \nonumber\\
                             &  \left.+(x_1-a_1)\arctan\sqrt{-\frac{a_1}{x_1}}-(x_1-a_2)\arctan\sqrt{-\frac{a_2}{x_1}}\right],
                             \label{uloggamma}
 \end{align}
 \begin{align}
 \langle \Gs_{21} \rangle^{(+)}(x_1) & = -\cfrac{\theta_s \varXi^+_t}{2 b}\bigg(1-\frac{\gamma_t^-}{\gamma_t^+}\bigg)[H(a_1-x_1)-H(a_2-x_1)],        \\
 \langle \Gs_{22} \rangle^{(+)}(x_1) & = -\cfrac{1}{b}\mS^{(c)}\cfrac{\partial \jump{0.1}{u_2}^{(-)}}{\partial x_1}+\cfrac{\theta_s  \varXi^+_t}{4\pi b}\bigg(1+\frac{\gamma_t^-}{\gamma_t^+}\bigg)\left\{\ln\left[(x_1-a_1)^2\right] 
 -\ln\left[(x_1-a_2)^2\right]\right\}.
\label{sigmaloggamma}
\end{align}
Note that, as expected, for the case of homogeneous material, corresponding to $\gamma_t^+=\gamma_t^-$, $\langle \Gs_{21} \rangle^{(+)}$ vanishes indipendently of 
the results derived by applying the integral operators to the temperature profiles. This means that for the case of a plane crack in an homogeneous elastic themodiffusive material, 
no shear stresses are due to the temperature and heat flux profiles on the crack surface.

\begin{figure}[!htcb]
\centering
\includegraphics[width=13.6cm]{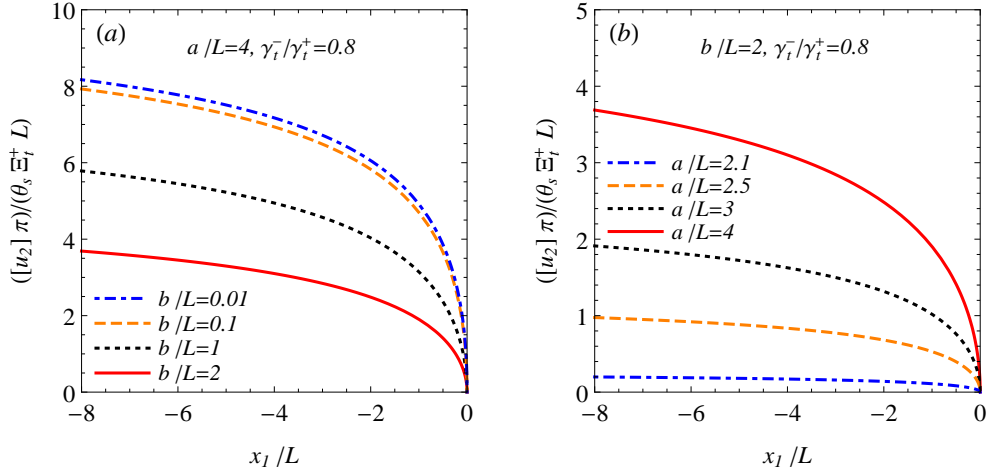}
\caption{\footnotesize (a): Variation of the normalized crack opening $\jump{0.1}{u_2}^{(-)}$ with the spatial coordinate $x_1/L$, reported for different values
of $a_1/L$;  (b):  Variation of the normalized crack opening $\jump{0.1}{u_2}^{(-)}$ with the spatial coordinate $x_1/L$, reported for different values
of $a_2/L$.}
\label{figu2jumplog}
\end{figure}

The variation of the normalized crack opening \eq{uloggamma} is reported in Fig. \ref{figu2jumplog} as a function of the spatial coordinate $x_1/L$ assuming a single value of the ratio
$\gamma_t^-/\gamma_t^+=0.8$ and different localization of the Dirac Delta heat flux functions \eq{qlog1} and \eq{qlog2}, corresponding to different values of the normalized distances $a_1/L$ and
$a_2/L$. The values for these distances have been chosen such that $a_1>a_2$, as imposed in the definition of the heat flux profiles \eq{qlog1} and \eq{qlog2}.
The crack opening profiles shown in Fig. \ref{figu2jumplog}$/(a)$ have been computed considering the same value of $a_1/L=0.8$ and four different values of $a_2/L=\{0.001, 0.1, 1,2\}$.
It can be observed that, as $a_2/L$ decreases, and then the contribution to the heat flux functions depending by $\delta((x_1-a_2)/L)$ approaches the crack tip, the crack opening increases. 
In the author's opinion, this is due to the fact that, for small values of $a_2/L$, the peak of the heat flux profiles \eq{qlog1} and \eq{qlog2} 
and the maximum of the temperature \eq{tlog} are localized near to the crack tip. Consequently the maximum of the thermal load provided by these distributions of heat flux and temperature
approaches the crack tip, the crack process is favored and an increase of the crack opening is produced, in analogy to what is observed for mechanical loading functions localized 
near to the tip \citep{freund1}.

In Fig. \ref{figu2jumplog}$/(b)$ the crack opening $\jump{0.1}{u_2}^{(-)}$  is reported for the same value of $a_2/L=2$ and four different values of $a_1/L=\{2.1, 2.5, 3, 4\}$. Differently
from what is detected for the variation of $a_2/L$, it can be noted that the crack opening becomes larger as the normalized distance $a_1/L$ increases. As we expect, $\jump{0.1}{u_2}^{(-)}$ is almost
zero for $a_1/L=2.1$, since in this case $a_1\approx a_2$, and then observing expressions \eq{qlog1}, \eq{qlog2} and \eq{tlog} it can be easily deduced that $\left \langle q_1 \right\rangle\approx 0$, 
$\jump{0.1}{q_1}\approx 0$ and $\left\langle \theta \right\rangle\approx 0$. Looking at the structure of profiles \eq{qlog1}, \eq{qlog2} and  \eq{tlog}, it can be easily deduced that if 
the value of $a_2/L$ is fixed and $a_1/L$ increase, the effort of the temperature and heat fluxes becomes more relevant. Consequently, the loading action provided by the temperature 
and the flux at the interface increases and gives rise to higher values of the crack opening. 

\begin{figure}[!htcb]
\centering
\includegraphics[width=13.6cm]{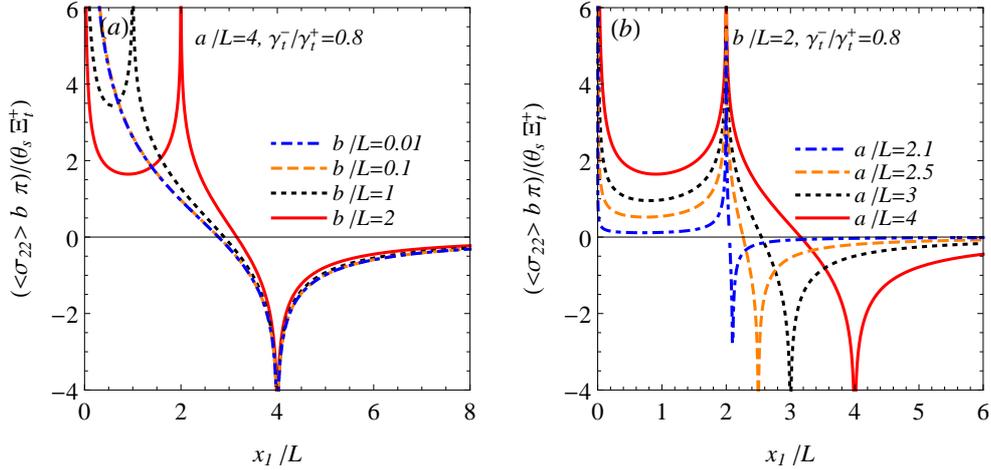}
\caption{\footnotesize  (a): Variation of the normalized traction $\langle \Gs_{22} \rangle^{(+)}$ with the spatial coordinate $x_1/L$, reported for different values
of $a_1/L$;  (b):  Variation of the normalized traction $\langle \Gs_{22} \rangle^{(+)}$ with the spatial coordinate $x_1/L$, reported for different values
of $a_2/L$.}
\label{figsigma22log}
\end{figure}

The variation of the normalized traction $\langle \Gs_{22} \rangle^{(+)}$ is plotted in  Fig. \ref{figsigma22log} as a function of the spatial coordinate $x_1/L$ 
for the same single value of the ratio $\gamma_t^-/\gamma_t^+=0.8$ and the same sets of values of $a_1/L$ and $a_2/L$ assumed for the crack opening. In order to compute profiles of the traction $\langle \Gs_{22} \rangle^{(+)}$,
the symbolic computation program Mathematica was used. As it can be expected, the tractions exhibit a singular 
behaviour at the crack tip, for $x_1=0$, and for $x_1=a_1$ and $x_1=a_2$, where the temperature profile \eq{tlog} diverges.


\begin{figure}[!htcb]
\centering
\includegraphics[width=13.6cm]{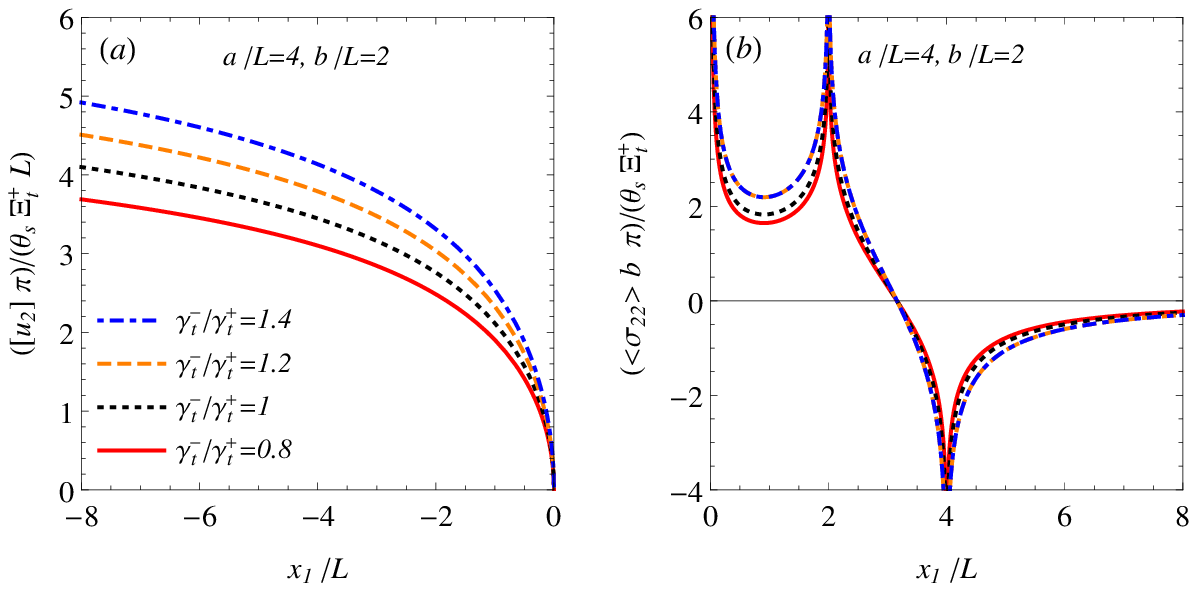}
\caption{\footnotesize  (a): Variation of the normalized crack opening $\jump{0.1}{u_2}^{(-)}$ with the spatial coordinate $x_1/L$, reported for different values
of the ratio $\gamma_t^-/\gamma_t^+$;  (b):  Variation of the normalized traction $\langle \Gs_{22} \rangle^{(+)}$ with the spatial coordinate $x_1/L$, reported for different values
of the ratio $\gamma_t^-/\gamma_t^+$.}
\label{figlogGamma}
\end{figure}

In Fig. \ref{figlogGamma} the crack opening $\jump{0.1}{u_2}^{(-)}$ and the traction ahead of the tip are reported as functions of $x_1/L$ for single values of $a_1/L=4$ and of $a_2/L=2$ and several
values of the ratio $\gamma_t^-/\gamma_t^+=\{0.8, 1.2, 1, 0.8 \}$. It can be noted that greater values of $\gamma_t^-/\gamma_t^+$ correspond to a greater crack opening. This means that,
taking into account the considered heat flux, the temperature profiles and the geometry of the model, the crack process is favored if the thermal expansion coefficient of the material occupying the
lower half-space is greater with respect to that of the medium occupying the upper half-plane.

\begin{figure}[!htcb]
\centering
\includegraphics[width=14.8cm]{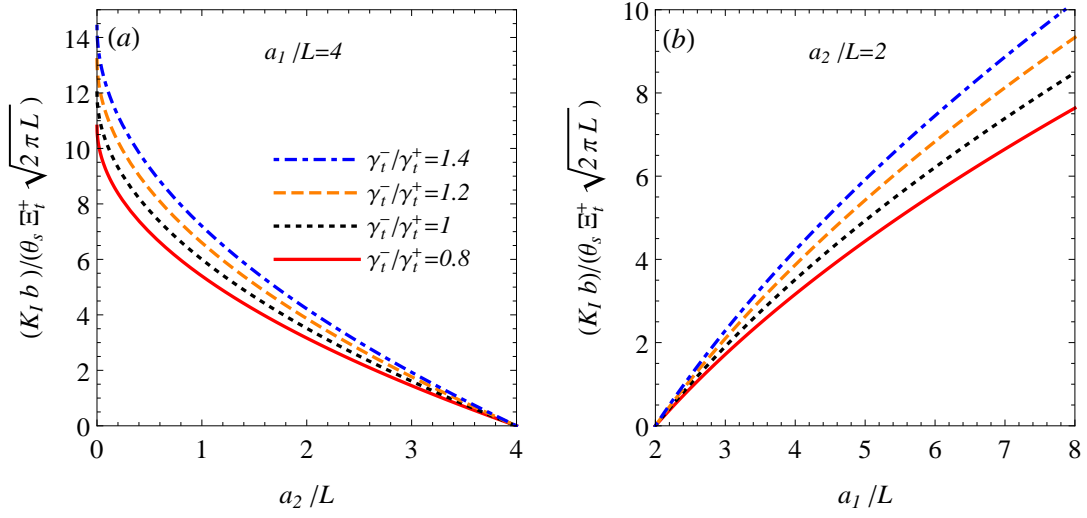}
\caption{\footnotesize  (a): Variation of the normalized stress intensity factor $K_{I}$  with the ratio $a_2/L$, reported for $a_{1}/L=2$ and assuming different values
of $\gamma_t^-/\gamma_t^+$;  (b):  Variation of the normalized stress intensity factor $K_{I}$ with the ratio $a_1/L$, reported for $a_{2}/L=2$ and assuming different values
of the ratio $\gamma_t^-/\gamma_t^+$.}
\label{figlogK}
\end{figure}

The traction expressions \eq{sigmaloggamma} can now be used for calculating the stress intensity factors. Applying the standard definition reported 
in \citet{freund1}, we obtain
\begin{align}
 K_{I} & =\lim_{x_1\rightarrow 0}\sqrt{2 \pi x_1}\langle \Gs_{22} \rangle^{(+)}(x_1)=\frac{3\sqrt{2\pi}\theta_s \varXi^+_t}{b}
 \bigg(1+\frac{\gamma_t^-}{\gamma_t^+}\bigg)\bigg(\sqrt{\frac{a_1}{L}}-\sqrt{\frac{a_2}{L}}\bigg), \label{KIlog} \\
 K_{II}& =\lim_{x_1\rightarrow 0}\sqrt{2 \pi x_1}\langle \Gs_{21} \rangle^{(+)}(x_1)=0.
\end{align}
In Fig. \ref{figlogK}$/(a)$ the variation of the normalized stress intensity factor $K_{I}$ with the ratio $a_{2}/L$ is reported for $a_{1}/L=4$ and several 
values the ratio $\gamma_t^-/\gamma_t^+=\{0.8, 1.2, 1, 0.8 \}$. It can be observed that $K_{I}=0$ for $a_{2}/L=a_{1}/L=4$. 
This is due to the fact that for  $a_{2}/L=a_{1}/L$ both the heat flux and temperature profiles vanish (see expressions \eq{qlog1}, \eq{qlog2} and \eq{tlog}), and 
consequently no thermal load is present at the interface and no crack opening is generated. Moreover, it can be noted that as $a_{2}/L$ decreases, $K_{I}$ increases monotonically. This
behaviour is in agreement with what is detected analysing the crack opening, and also in this case it can be explained observing that for small values of $a_{2}/L$ 
the heat flux profiles \eq{qlog1} and \eq{qlog2} and the temperature \eq{tlog} possess a peak localized at the crack tip. As a consequence, the maximum of the provided thermal load 
approaches the crack tip and the crack process is favored. In Fig. \ref{figlogK}$/(b)$, $K_{I}$ is plotted as a function of $a_{1}/L$ assuming 
 $a_{2}/L=2$ and the same values $\gamma_t^-/\gamma_t^+=\{0.8, 1.2, 1, 0.8 \}$. According to what is observed in  Fig. \ref{figlogK}$/(a)$,  $K_{I}=0$ for $a_{2}/L=a_{1}/L=2$.
 The reported curves show that the stress intensity factor becomes larger as $a_{1}/L$ increases. This behaviour is due to the characteristics 
 of the heat flux and temperature expressions \eq{qlog1}, \eq{qlog2} and \eq{tlog}: looking at the structure of these profiles, it can be easily verified 
 that if the value of $a_{2}/L$ is fixed and$a_{1}/L$, the effort of temperature and heat fluxes becomes more relevant, then the loading action provided by thermal effects increases.
 Consequently, the crack process is favored and the value of $K_{I}$ becomes larger as does the crack opening (see Fig. \ref{figu2jumplog}$/(b)$). Observing both Fig. \ref{figlogK}$/(a)$
 and Fig. \ref{figlogK}$/(b)$, it can be noted that the stress intensity factor increases as the ratio $\gamma_t^-/\gamma_t^+$. This confirms the behaviour detected observing the crack opening,
 and it means that considering the heat flux and temperature profiles given by expressions \eq{qlog1}, \eq{qlog2} and \eq{tlog}, 
 the crack process is favored if the thermal expansion coefficient of the material occupying the lower half-plane is greater with respect to that of the medium occupying the upper half-plane.

Considering the case of a homogeneous material, where we have $\gamma_t^-=\gamma_t^+=\gamma_t$, the crack opening \eq{uloggamma} assume the form
 \begin{align}
  \jump{0.1}{u_1}^{(-)}(x_1) & = 0,               \\   
  \jump{0.1}{u_2}^{(-)}(x_1) & =  \frac{\gamma_t\theta_s}{\pi(\lambda+\mu)}\Bigg[\frac{\pi}{2}(a_1-a_2)+(\sqrt{-a_1x_1}-\sqrt{-a_2x_1}) \nonumber\\
                        &  \left.+(x_1-a_1)\arctan\sqrt{-\frac{a_1}{x_1}}-(x_1-a_2)\arctan\sqrt{-\frac{a_2}{x_1}}\right],\label{u2hom}
 \end{align}
 and then the tractions ahead of the tip \eq{sigmaloggamma} become
 \begin{align}
 \langle \Gs_{21} \rangle^{(+)}(x_1) & = 0,         \\
 \langle \Gs_{22} \rangle^{(+)}(x_1) & = -\cfrac{1}{b}\mS^{(c)}\cfrac{\partial \jump{0.1}{u_2}^{(-)}}{\partial x_1}
 +\cfrac{\mu\gamma_t\theta_s}{4\pi(\lambda+2\mu)}\left\{\ln\left[(x_1-a_1)^2\right] -\ln\left[(x_1-a_2)^2\right]\right\}, \label{sigma22hom}
\end{align}
where 
\begin{equation}
 \cfrac{1}{b}\frac{\partial \jump{0.1}{u_2}^{(-)}}{\partial x_1}=\cfrac{\mu\gamma_t\theta_s}{\pi(\lambda+2\mu)}
 \left[-\sqrt{-\frac{a_1}{x_1}}+\sqrt{-\frac{a_2}{x_1}}+\arctan\sqrt{-\frac{a_1}{x_1}}-\arctan\sqrt{-\frac{a_2}{x_1}}\right]. \label{u2dhom}
\end{equation}
Expressions \eq{u2hom}, \eq{sigma22hom} and \eq{u2dhom} can be easily obtained deriving the Green's function corresponding to a thermoelastic semi-plane with the punctually localized fluxes 
\eq{qlog1} and \eq{qlog2} applied on the boundary.

\subsection{Symmetrically distributed temperature profile at the interface}
\begin{figure}[!htcb]
\centering
\includegraphics[width=13.6cm]{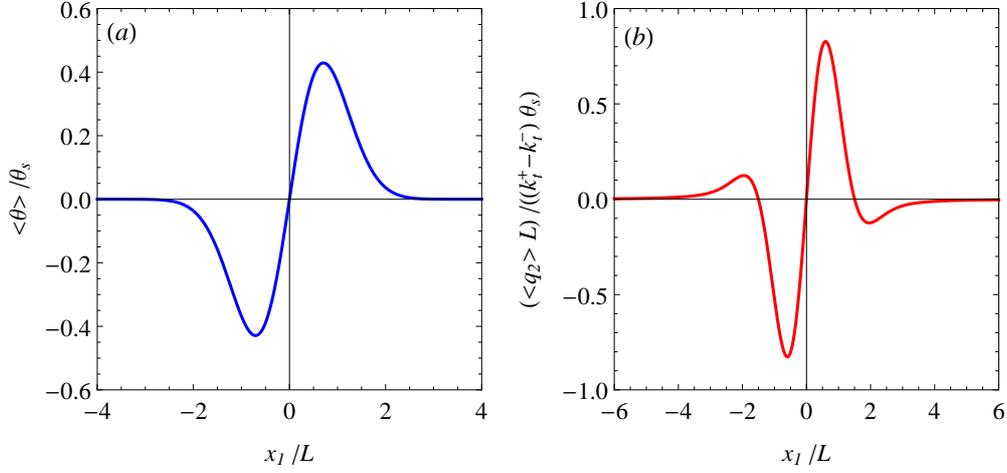}
\caption{\footnotesize (a): Normalized temperature profile at the interface $x_2=0$; (b): Normalized average heat flux profile at the interfaces $x_2=0$.}
\label{figthetaq}
\end{figure}

The following distribution profile for the avegare temperature across the plane $x_2=0$, containing both the crack and the interface, is considered:
\beq
\left\langle \theta \right\rangle (x_1)=\frac{\theta_s x_1}{L}e^{-\frac{x_1^2}{L^2}}.
\label{thetadist}
\eeq
The temperature is assumed to be continous at the interface, then $\jump{0.1}{\theta}(x_1)=0$ for $-\infty<x_1<+\infty$. The average and the jump of the normal heat flux $q_2$ at the interface
corresponding to the average temperature profile \eq{thetadist} and to the assumption $\jump{0.1}{\theta}(x_1)=0$ has been evaluated solving the steady state 
 heat conduction equation for both the upper and the lower thermodiffusive half-space. This procedure is reported details in Appendix \ref{thetagauss}, and the resulting average and jump of the
 of the noramal heat flux are given by
\beq
\left\langle q_2 \right\rangle (x_1)=-(k_t^+ -k_t^-)\frac{\theta_s}{L}\left[\left
(\frac{2 x_1^2}{L^2}-1\right)e^{-\frac{x_1^2}{L^2}}\mbox{Erfi}\left(\frac{x_1}{L}\right)-\frac{2 x_1}{L\sqrt{\pi}}\right],
\label{qavdist}
\eeq
\beq
\jump{0.1}{q_1}(x_1)=-2(k_t^+ +k_t^-)\frac{\theta_s}{L}\left[\left
(\frac{2 x_1^2}{L^2}-1\right)e^{-\frac{x_1^2}{L^2}}\mbox{Erfi}\left(\frac{x_1}{L}\right)-\frac{2 x_1}{L\sqrt{\pi}}\right].
\label{qjumpdist}
\eeq
It can be easily verified that the normal heat flux profiles \eq{qavdist} and \eq{qjumpdist} satisfy the integral balance conditions introduced in Section \ref{intercrack}. The variation
of the normalized average temperature \eq{thetadist} and of the normalized average heat flux \eq{qavdist} is plotted as a function of the spatial coordinate $x_1/L$ in Fig. \ref{figthetaq}.

Also for this example, the Dundurs parameter $d$ involved in the weight functions \eq{weight_expl} is assumed to be zero, and then considering the contributions
due to the fluxes and temperature profiles, the uncoupled integral identities have the form \eq{id-log}. Applying the integral operator $\mJ^{(s-)}$ to the average and jump of the normal heat flux
\eq{qavdist} and \eq{qjumpdist}, the following result is obtained
\begin{equation}
 \frac{h_t}{2}\mJ^{(s-)}\langle q_2 \rangle+\frac{l_t}{4}\mJ^{(s-)}\jump{0.1}{ q_2}=\frac{\theta_s\varUpsilon_t x_1}{2L}e^{-\frac{x_1^2}{L^2}}.
\label{Js-}
 \end{equation}
Substituting expression \eq{Js-} into equations \eq{id-log}, we get
\begin{equation}
\left\{
\begin{array}{rrll}
 \mS^{(s)}\cfrac{\partial \jump{0.1}{u_1}^{(-)}}{\partial x_1} & = &  m_t\theta_s\mS^{(s-)}\bigg(\cfrac{x_1}{L}e^{-\frac{x_1^2}{L^2}}\bigg),  &        \\
                                                               &   &                                     &  x_1<0\\
 \mS^{(s)}\cfrac{\partial \jump{0.1}{u_2}^{(-)}}{\partial x_1} & = & -\theta_s\bigg(n_t-\cfrac{\varUpsilon_t}{2}\bigg)\cfrac{x_1}{L}e^{-\frac{x_1^2}{L^2}}
\end{array}
\right.,
\end{equation}
where $\varUpsilon_t$ is the same bimaterial parameter defined in Section \ref{exlog}.
Using the procedure reported in \citet{PiccMish3} and \citet{MorPicc2}, the singular integral operator $\mS^{(s)}$ 
can be inverted. This leads to
\begin{align}
  \frac{\partial \jump{0.1}{u_1}^{(-)}}{\partial x_1} & =  m_t\theta_s\mS^{(s)-1}\mS^{(s-)}\bigg(\cfrac{x_1}{L}e^{-\frac{x_1^2}{L^2}}\bigg),         \label{du1}      \\   
  \frac{\partial \jump{0.1}{u_2}^{(-)}}{\partial x_1} & =  -\theta_s\bigg(n_t-\cfrac{\varUpsilon_t}{2}\bigg)\mS^{(s)-1}\bigg(\cfrac{x_1}{L}e^{-\frac{x_1^2}{L^2}}\bigg).  \label{du2}                      
 \end{align}
 
 The results provided by the application of the operators $\mS^{(s)-1}$ and $\mS^{(s-)}$ to the function $(x_1/L)\cdot e^{-\frac{x_1^2}{L^2}}$
 are calculated by means of the symbolic computation program Mathematica. The crack opening $\jump{0.1}{u_1}^{(-)}$ and
 $\jump{0.1}{u_2}^{(-)}$ are derived by integrating respectively expressions \eq{du1} and \eq{du2}.
 
 Applying the operators  $\mS^{(s+)}$ to the flux functions \eq{qavdist} and \eq{qjumpdist}, the following result is derived
 \begin{equation}
 \frac{h_t}{2}\mJ^{(s+)}\langle q_2 \rangle+\frac{l_t}{4}\mJ^{(s+)}\jump{0.1}{ q_2}=\frac{\theta_s\Upsilon_t x_1}{2L}e^{-\frac{x_1^2}{L^2}}.
\label{Js+}
 \end{equation}
Substituting expression \eq{Js+} into equations \eq{id2log}, the tractions ahead of the crack tip become
 \begin{align}
 \langle \Gs_{21} \rangle^{(+)}(x_1) & = \cfrac{m_t\theta_s}{b}\bigg[\mS^{(s+)}\bigg(\cfrac{x_1}{L}e^{-\frac{x_1^2}{L^2}}\bigg)
 -\mS^{(c)}\mS^{(s)-1}\mS^{(s-)}\bigg(\cfrac{x_1}{L}e^{-\frac{x_1^2}{L^2}}\bigg)\bigg],    \label{sig21}     \\
 \langle \Gs_{22} \rangle^{(+)}(x_1) & = -\cfrac{\theta_s}{b}\bigg(n_t-\cfrac{\varUpsilon_t}{2}\bigg)\bigg[\cfrac{x_1}{L}e^{-\frac{x_1^2}{L^2}}
 -\mS^{(c)}\mS^{(s)-1}\bigg(\cfrac{x_1}{L}e^{-\frac{x_1^2}{L^2}}\bigg)\bigg].\label{sig22}
\end{align}

Also for the example described in this Section, the upper and lower half-space media are assumed to have identical values for the 
 elastic moduli $\lambda^+=\lambda^-=\lambda$, $\mu^+=\mu^-=\mu$, and for the thermal conductivity coefficients $k^+_t=k^-_t=k_t$, and dissimilar values for $\gamma_t$:
$\gamma_t^+\neq \gamma_t^-$. In this case, the equations \eq{du1} and \eq{du2} become
\begin{align}
  \frac{\partial \jump{0.1}{u_1}^{(-)}}{\partial x_1} & =  \theta_s\varXi_t^+\bigg(1-\frac{\gamma_t^-}{\gamma_t^+}\bigg)\mS^{(s)-1}\mS^{(s-)}\bigg(\cfrac{x_1}{L}e^{-\frac{x_1^2}{L^2}}\bigg), \label{du1dist}              \\   
  \frac{\partial \jump{0.1}{u_2}^{(-)}}{\partial x_1} & =  -\theta_s\varXi_t^+\bigg(1+\frac{\gamma_t^-}{\gamma_t^+}\bigg)\mS^{(s)-1}\bigg(\cfrac{x_1}{L}e^{-\frac{x_1^2}{L^2}}\bigg),  \label{du2dist}                      
 \end{align}
 and the expressions \eq{sig21} and \eq{sig22} assume the form
 \begin{align}
 \langle \Gs_{21} \rangle^{(+)}(x_1) & = \cfrac{\theta_s \varXi_t^+ }{b}\bigg(1-\frac{\gamma_t^-}{\gamma_t^+}\bigg)\bigg[\mS^{(s+)}\bigg(\cfrac{x_1}{L}e^{-\frac{x_1^2}{L^2}}\bigg)
 -\mS^{(c)}\mS^{(s)-1}\mS^{(s-)}\bigg(\cfrac{x_1}{L}e^{-\frac{x_1^2}{L^2}}\bigg)\bigg],  \label{sig21dist}       \\
 \langle \Gs_{22} \rangle^{(+)}(x_1) & = -\cfrac{\theta_s \varXi_t^+}{b}\bigg(1+\frac{\gamma_t^-}{\gamma_t^+}\bigg)\bigg[\cfrac{x_1}{L}e^{-\frac{x_1^2}{L^2}}
 -\mS^{(c)}\mS^{(s)-1}\bigg(\cfrac{x_1}{L}e^{-\frac{x_1^2}{L^2}}\bigg)\bigg], \label{sig22dist}
\end{align}
where the quantity $\varXi_t^+$ is the same defined in Section \ref{exlog}.

\begin{figure}[!htcb]
\centering
\includegraphics[width=13.6cm]{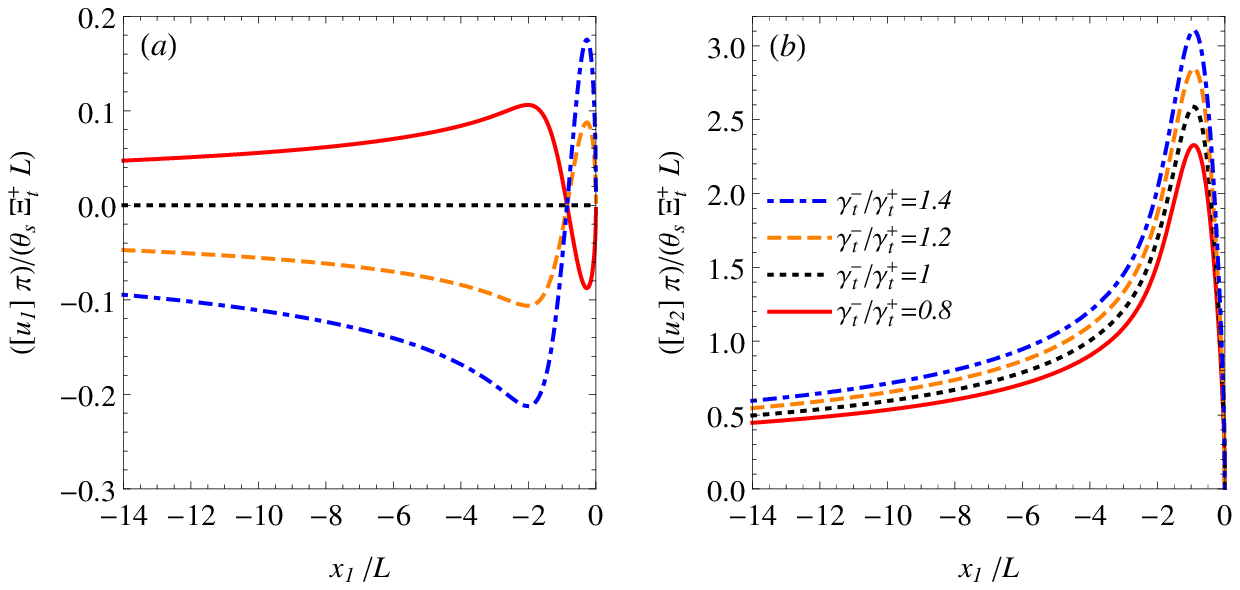}
\caption{\footnotesize (a): Variation of the normalized crack opening $\jump{0.1}{u_1}^{(-)}$ with the spatial coordinate $x_1/L$, reported for different values of the ratio $\gamma_t^-/\gamma_t^+$;
(b):  Variation of the normalized crack opening $\jump{0.1}{u_2}^{(-)}$ with the spatial coordinate $x_1/L$,  reported for different values of the ratio $\gamma_t^-/\gamma_t^+$.}
\label{figujump}
\end{figure}
\begin{figure}[!htcb]
\centering
\includegraphics[width=13.6cm]{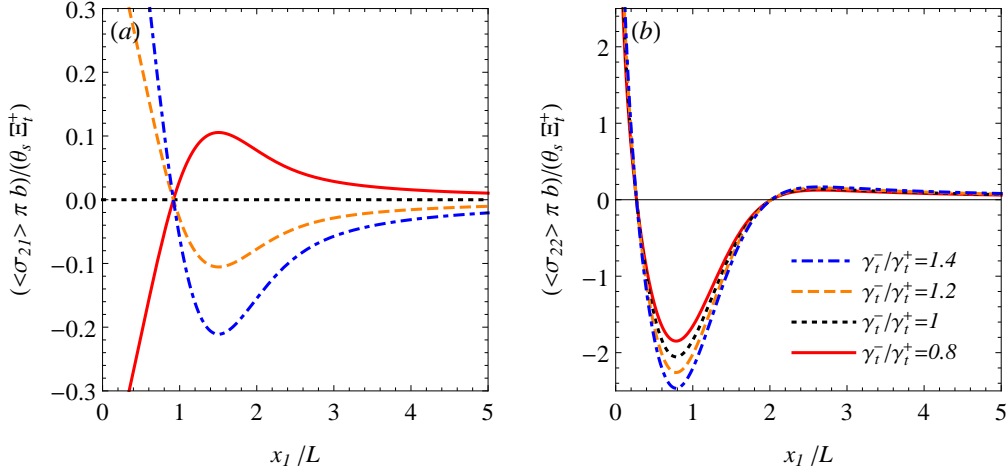}
\caption{\footnotesize (a): Variation of the normalized traction $\langle \Gs_{21} \rangle^{(+)}$ with the spatial coordinate $x_1/L$, reported for different values
of the ratio $\gamma_t^-/\gamma_t^+$;  (b):  Variation of the normalized crack traction $\langle \Gs_{22} \rangle^{(+)}$ with the spatial coordinate $x_1/L$, reported for different values
of the ratio $\gamma_t^-/\gamma_t^+$.}
\label{figsigma}
\end{figure}

The normalized profiles of the crack opening components $\jump{0.1}{u_1}^{(-)}$ and $\jump{0.1}{u_2}^{(-)}$, obtained by integrating expressions \eq{du1dist} and \eq{du2dist}, are reported in Fig. \ref{figujump}
as function of $x_1/L$. Four different values of the ratio $\gamma_t^-/\gamma_t^+=\{0.8, 1.2, 1, 0.8 \}$ were considered for the computations. For $\gamma_t^-/\gamma_t^+=1$, corresponding to
the case of an homogeneous material, we have $\jump{0.1}{u_1}^{(-)}=0$. This behaviour can be directly deduced observing expression \eq{du1dist}. Observing  Fig. \ref{figujump}$/(a)$, it can 
be noted that the crack opening component $\jump{0.1}{u_1}^{(-)}$ changes sign at a determinate distance $x_1^{*}/L$ from the crack tip. As it is shown in the figure, this distance is independent of the 
ratio $\gamma_t^-/\gamma_t^+$, it then varies with the value of $a/L$ and $b/L$. For $\gamma_t^-/\gamma_t^+>1$, $\jump{0.1}{u_1}^{(-)}$ is positive near to the crack tip, and then becomes
negative for $x_1/L<x_1^{*}/L$. This means that, if $\gamma_t^->\gamma_t^+$, in the $x_1-$direction, associated with the propagation Mode II, the crack opens for $x_1/L>x_1^{*}/L$ and closes for $x_1/L<x_1^{*}/L$. 
The opposite behaviour is shown for $\gamma_t^-/\gamma_t^+<1$, corresponding to a crack opening for $x_1/L<x_1^{*}/L$ and to a crack closing for $x_1/L>x_1^{*}/L$, near to the tip.
Conversely, the component  $\jump{0.1}{u_2}^{(-)}$ illustrated in Fig. \ref{figujump}$/(b)$ is positive everywhere for any value of $\gamma_t^-/\gamma_t^+$. It follows that in
the $x_2-$direction, corresponding to the propagation Mode I we have crack opening for any value of the thermal expansion coefficients of the materials.

The variation of the normalized tractions $\langle \Gs_{21} \rangle^{(+)}$ and $\langle \Gs_{22} \rangle^{(+)}$ is plotted in  Fig. \ref{figsigma} as a function of the spatial coordinate $x_1/L$ 
for the same values  of $\gamma_t^-/\gamma_t^+$ assumed for the crack opening. In order to compute the integrals involved in equations \eq{sig21dist} and \eq{sig22dist}
the symbolic computation program Mathematica was used. It can be observed that both $\langle \Gs_{21} \rangle^{(+)}$ and $\langle \Gs_{22} \rangle^{(+)}$ exhibit a singular 
behaviour at the crack tip, for $x_1=0$. This behaviour is associated with the leading term of the asymptotic expansion of expressions  \eq{sig21dist} and \eq{sig22dist} for $x_1/L\rightarrow 0$,
corresponding respectively to the thermal stress intensity factors $K_t^{II}$ and $K_t^{I}$ \citep{LiuKar1}.

Also in this case, exact expressions for the stress intensity factors can be evaluated by means of the tractions \eq{sig21dist} and \eq{sig22dist}.
Applying the standard definition reported in \citet{freund1}, we obtain
\begin{align}
 K_{I} & =\lim_{x_1\rightarrow 0}\sqrt{2 \pi x_1}\langle \Gs_{22} \rangle^{(+)}(x_1)=\frac{3\pi^{3/2}\sqrt{L}\theta_s \varXi^+_t}{4b}
 \bigg(1+\frac{\gamma_t^-}{\gamma_t^+}\bigg)\left(\frac{\Gamma(3/4)}{\Gamma(1/4)\Gamma(7/4)}\right), \label{KI} \\
 K_{II}& =\lim_{x_1\rightarrow 0}\sqrt{2 \pi x_1}\langle \Gs_{21} \rangle^{(+)}(x_1)=-\frac{\pi^{3/2}\sqrt{L}\theta_s \varXi^+_t}{b\Gamma(1/4)}
 \bigg(1-\frac{\gamma_t^-}{\gamma_t^+}\bigg).\label{KII}
\end{align}
\section{Conclusions}
The problem of a quasi-static semi-infinite interfacial crack between two dissimilar thermodiffusive elastic materials has been formulated in terms of
boundary integral equations by means of Betti's reciprocity identity and weight functions. For the case of a plane strain crack, explicit integral identities relating
the applied mechanical loading, the temperature, mass density, heat and mass flux distributions at the interface and the resulting crack opening have been obtained. The solution
of this system of singular integral equations provides explicit expressions for the crack opening and the stress fields associate to any arbitrary harmonic temperature and 
mass density profile and self-balanced heat or mass flux distribution at the interface.

Illustrative examples of the application of the integral identities to crack problems characterized by punctually localized and distributed heat flux profiles on the interface 
were performed. The crack opening and tractions ahead of the tip corresponding to the assumed temperature and heat flux distributions were derived by analytical inversion
of the singular integral operators. In addition, exact expressions for the stress intensity factors were obtained.

The proposed original method provides a general boundary integral formulation for interfacial crack problems in thermodiffusive bimaterials avoiding the use of the Green's function
and the related challenging numerical calculations. The integral identities can be applied for studying the effects of any harmonic temperature and mass concentration profile
as well as of any self-balanced heat and mass flux distributions on crack opening and traction fields ahead of the tip. This approach can have various relevant applications,
expecially in the modelling of fracture processes induced by thermal and diffusive stresses at the interface between different components of mutlti-layered electrochemical devices such as solid oxide
fuel cells and lithium ions batteries. Moreover, the derived integral identities have their own value from the mathematical point of view, because, 
to the authors best knowledge, a similar explicit formulation in terms of singular integral equations seems to be unknown in the literature.

\section*{Acknowledgments}
The authors gratefully acknowledge financial support from the Italian Ministry of Education,
University and Research in the framework of the FIRB project 2010 "Structural mechanics models for renewable energy applications".

\newpage

\appendix
\newpage
\section{Bimaterial parameters}
In this Appendix we report explicit expressions for the bimaterial parameters defined in Sections 2 and 3. The parameters $b, d, \gamma$ and $\alpha$ contained in the weight functions 
matrices \eq{weight_expl} are the same as defined in \citet{PiccMish1}, here expressed in function of $\lambda, \mu$ instead of $\nu, \mu$:
\beq
b=\frac{\lambda^++2\mu^+}{2\mu^+(\lambda^++\mu^+)}+\frac{\lambda^-+2\mu^-}{2\mu^-(\lambda^-+\mu^-)}, \quad d=\frac{1}{2(\lambda^++\mu^+)}-\frac{1}{2(\lambda^-+\mu^-)},
\eeq
\beq
\alpha=\frac{\mu^-(\lambda^-+\mu^-)(\lambda^++2\mu^+)-\mu^+(\lambda^++\lambda^+)(\lambda^-+2\mu^-)}{\mu^-(\lambda^-+\mu^-)(\lambda^++2\mu^+)+\mu^+(\lambda^++\lambda^+)(\lambda^-+2\mu^-)},
\eeq
\beq
\gamma=\frac{\mu^+\mu^-(\lambda^-+\mu^-)+\mu^+\mu^-(\lambda^++\mu^+)}{\mu^-(\lambda^++2\mu^+)(\lambda^-+\mu^-)+\mu^+(\lambda^-+2\mu^-)(\lambda^++\mu^+)}.
\eeq

The bimaterial parameters associated with the effects of the temperature and heat diffusion are given by: 
\beq
h_t=\frac{\gamma_t^+}{4k_t^+\mu^+}-\frac{\gamma_t^-}{4k_t^-\mu^-}, 
\quad \ell_t=\frac{\gamma_t^+}{4k_t^+\mu^+}+\frac{\gamma_t^-}{4k_t^-\mu^-},
\eeq
\beq
m_t=\frac{\gamma_t^+}{2(\lambda^++\mu^+)}-\frac{\gamma_t^-}{2(\lambda^-+\mu^-)}, 
\quad n_t=\frac{\gamma_t^+(\lambda^++3\mu^+)}{4\mu^+(\lambda^++\mu^+)}+\frac{\gamma_t^-(\lambda^-+3\mu^-)}{4\mu^-(\lambda^-+\mu^-)},
\eeq
\beq
p_t=\frac{\gamma_t^+}{2(\lambda^++\mu^+)}+\frac{\gamma_t^-}{2(\lambda^-+\mu^-)}, 
\quad q_t=\frac{\gamma_t^+(\lambda^++3\mu^+)}{4\mu^+(\lambda^++\mu^+)}-\frac{\gamma_t^-(\lambda^-+3\mu^-)}{4\mu^-(\lambda^-+\mu^-)},
\eeq
and the constant corresponding to the concentration and the mass diffusion presents the following form:
\beq
h_c=\frac{\gamma_c^+}{4D_c^+\mu^+}-\frac{\gamma_c^-}{4D_c^-\mu^-}, 
\quad \ell_c=\frac{\gamma_c^+}{4D_c^+\mu^+}+\frac{\gamma_c^-}{4D_c^-\mu^-},
\eeq
\beq
m_c=\frac{\gamma_c^+}{2(\lambda^++\mu^+)}-\frac{\gamma_c^-}{2(\lambda^-+\mu^-)}, 
\quad n_c=\frac{\gamma_c^+(\lambda^++3\mu^+)}{4\mu^+(\lambda^++\mu^+)}+\frac{\gamma_c^-(\lambda^-+3\mu^-)}{4\mu^-(\lambda^-+\mu^-)},
\eeq
\beq
p_c=\frac{\gamma_c^+}{2(\lambda^++\mu^+)}+\frac{\gamma_c^-}{2(\lambda^-+\mu^-)}, 
\quad q_c=\frac{\gamma_c^+(\lambda^++3\mu^+)}{4\mu^+(\lambda^++\mu^+)}-\frac{\gamma_c^-(\lambda^-+3\mu^-)}{4\mu^-(\lambda^-+\mu^-)}.
\eeq
\newpage
 \section{Elastic potentials in weight functions space}
 In this Appendix, the derivation of the explicit expressions for the Fourier transform of the elastic potentials \eq{PhiPsiexpl} is explained in details. These expressions are given by the
 solution of the ODEs system \eq{ODE}, which is shown here
 \beq
\left\{
\begin{array}{c}
\tilde{\varPsi}^{'}-i\xi\tilde{\varPhi}=\tilde{U}_{1}  \\  
\tilde{\varPhi}^{'}+i\xi\tilde{\varPsi}=\tilde{U}_{2}
\end{array}
\right .,
\label{ODE2}
\eeq
 where considering the upper half-plane $x_1>0$, the non-homgeneous terms in the system \eq{ODE2} are given by the weight functions 
 \begin{eqnarray}
 \tilde{U}_1(\xi,x_2) & = & \left\{\left[x_2-\frac{\lambda^{+} + 2\mu^+}{|\xi|(\lambda^{+}+\mu^+)}\right]\tilde{\varSigma}_{21}^{-}+
 i\left[\frac{\mu^+}{\xi(\lambda^{+}+\mu^{+})}-\sign(\xi)x_2 \right]\tilde{\varSigma}_{22}^{-}\right\}\frac{e^{-|\xi|x_2}}{2\mu^+},\nonumber\\
 \tilde{U}_2(\xi,x_2) & = & \left\{-i\left[\sign(\xi)x_2+\frac{\mu^+}{\xi(\lambda^{+}+\mu^{+})}\right]\tilde{\varSigma}_{21}^{-}-
 \left[x_2+\frac{\lambda^{+}+2\mu^{+}}{|\xi|(\lambda^{+}+\mu^{+})}\right]\tilde{\varSigma}_{22}^{-}\right\}\frac{e^{-|\xi|x_2}}{2\mu^+}.\nonumber\label{Uapp}\\
\end{eqnarray}

The ODEs system \eq{ODE2} can be rewritten in the matrix form:
\beq
\by^{'}+\bQ\by=\bc,
\label{ODEmatrix}
\eeq
where $\by=[\tilde{\varPsi}, \tilde{\varPhi}]^T$, $\bc=[\tilde{U}_{1}, \tilde{U}_{2}]^{T}$ and 
\beq
\bQ=\left[
\begin{array}{cc}
 0 &  -i\xi\\
 i\xi &  0
\end{array}
\right].
\label{Qmatrix}
\eeq
The eigenvectors and the eigenvalues of the matrix \eq{Qmatrix} are now defined:
\beq
(\bQ-\lambda\bI)\bw=0.
\eeq
The eigeinvalues are easily determined by solving the characteristic equation:
\beq
\mbox{det}(\bQ-\lambda\bI)=\lambda^2-\xi^2=0, \quad \Rightarrow \quad \lambda_1=|\xi|, \lambda_2=-|\xi|,
\eeq
then the eigenvector matrix $\bW=(\bw_1, \bw_2)^T$ and its inverse are given by 
\beq
\bW=\left[
\begin{array}{cc}
 i\sign(\xi) &  -i\sign(\xi)\\
 1 &  1
\end{array}
\right], \quad 
\bW^{-1}=\frac{1}{2}\left[
\begin{array}{cc}
 -i\sign(\xi) &  1\\
 i\sign(\xi) &  1
\end{array}
\right],
\label{Wmatrix}
\eeq
Applying the standard eigenvalues method and using the matrices \eq{Wmatrix} the system \eq{ODEmatrix} is written in the following equivalent form 
\beq
\bz^{'}+\bW^{-1}\bQ\bW\bz=\bd,
\label{ODEz}
\eeq
where $\bz=\bW^{-1}\by$, $\bd=\bW^{-1}\bc$ and 
\beq
\bW^{-1}\bQ\bW=\left[
\begin{array}{cc}
 |\xi| &  0\\
 0 &  -|\xi|
\end{array}
\right].
\label{WQmatrix}
\eeq
The \eq{ODEz} consists of two decoupled equations:
\beq
\left\{
\begin{array}{c}
z_{1}^{'}-|\xi|z_{1}=d_{1}  \\  
z_{2}^{'}+|\xi|z_{2}=d_{2}
\end{array}
\right .,
\label{uncoupled}
\eeq
where the non-homogeneous terms $\bd=\bW^{-1}\bc$  are given by
\beq
d_{1}=-\frac{1}{2}[i\sign(\xi)\tilde{U}_{1}-\tilde{U}_{2}], \quad d_{1}=\frac{1}{2}[i\sign(\xi)\tilde{U}_{1}+\tilde{U}_{2}].
\eeq

The general solutions of equations \eq{uncoupled} assume the form
 \begin{eqnarray}
 z_1(\xi,x_2) & = & C_1e^{|\xi|x_2}-\left[ix_2\tilde{\varSigma}_{21}^{-}+
 \left(\sign(\xi)x_2+\frac{1}{\xi}\right)\tilde{\varSigma}_{22}^{-}\right]\frac{e^{-|\xi|x_2}}{4\xi\mu^+},\nonumber\\
 z_2(\xi,x_2) & = & C_2e^{-|\xi|x_2}+\left(i\sign(\xi)\tilde{\varSigma}_{21}^{-}
 +\tilde{\varSigma}_{22}^{-}\right)\frac{(\lambda^+ + 3\mu^+)x_2 e^{-|\xi|x_2}}{4|\xi|(\lambda^+ + \mu^+)\mu^+}.\nonumber\label{zapp}\\
\end{eqnarray}
Assuming that the displacements and the stresses vanish for $x_1\rightarrow +\infty$, the following condition must be imposed: $C_1=0$. 
The expressions for the transformed elastic potentials $\by=\bW\bz=[\tilde{\varPsi}, \tilde{\varPhi}]^T$ in the upper half-plane can then be obtained by means of the 
following relations:
\beq
\tilde{\varPsi}=i\sign(\xi)(z_1-z_2), \quad \tilde{\varPhi}=z_1+z_2.
\label{z1z2}
\eeq
Substituting expressions \eq{zapp} in equations \eq{z1z2}, we get: 
\begin{eqnarray}
 \tilde{\varPhi}(\xi,x_2) & = &  C_2e^{-|\xi|x_2}+\left\{-i\frac{x_2\mu^+}{\xi(\lambda^{+}+\mu^+)}\tilde{\varSigma}_{21}^{-}+
 \left[\frac{1}{2\xi^2}-\frac{x_2\mu^+}{|\xi|(\lambda^{+}+\mu^{+})}\right]\tilde{\varSigma}_{22}^{-}\right\}\frac{e^{-|\xi|x_2}}{2\mu^+},\nonumber\\
 \tilde{\varPsi}(\xi,x_2) & = & i\sign(\xi)C_2e^{-|\xi|x_2}+\left\{-\frac{x_2(\lambda^{+}+2\mu^{+})}{|\xi|(\lambda^{+}+\mu^{+})}\tilde{\varSigma}_{21}^{-}
 +i\left[\frac{\sign(\xi)}{2\xi^2}+\frac{x_2(\lambda^{+}+2\mu^{+})}{\xi(\lambda^{+}+\mu^{+})}\right]\tilde{\varSigma}_{22}^{-}\right\}\frac{e^{-|\xi|x_2}}{2\mu^+}.\nonumber\\
\label{PhiPsiapp}
 \end{eqnarray}
assuming $C_2=0$, the expressions \eq{PhiPsiexpl} are finally derived. Similarly to the case of the displacements, 
replacing $-|\xi|$ with $|\xi|$, $\mu^+$ with $\mu^-$ and $\lambda^+$ with $\lambda^-$ in the \eq{PhiPsiapp} the elastic potentials for the lower half-plane are obtained.
\section{Temperature and heat flux profiles}
 The temperature and heat flux profiles used in Section \ref{examples} are now derived by solving the heat transmission problem in a semi-plane for different sets of boundary conditions.
 Considering the upper thermodiffusive half-plane $x_1>0$, the temperature profile $\theta^{+}(x_1,x_2)$ is given by the solution of the two-dimensional Laplace's equation 
 \beq
\Delta\theta^{+}=0, \quad x_2>0,
\label{harmapp}
\eeq
Two different sets of boundary conditions are considered for the solution of equation \eq{harmapp}. 
\subsection{Punctually localized heat flux profile at the interface}
\label{tlogapp}
The following profile for the temperature on the boundary $x_2=0^{+}$ is assumed
\beq
q_{2}^{+}(x_1, x_2=0^+)=\frac{\theta_s k_{t}^{+}}{2L}\left[\delta\left(\frac{x_1-a_1}{L}\right)-\delta\left(\frac{x_1-a_2}{L}\right)\right],
\label{qdelta}
\eeq
where $a_1, a_2, L >0$ and $a_1 > a_2$. It is easy to verify that the flux function \eq{qdelta} satisfies the self-balance condition \eq{balx1}$_{(1)}$.
Substituting expression \eq{qdelta} into the heat flux definition \eq{flux}$_{(1)}$ and remembering the condition \eq{fields_inf}$_{(1)}$, 
the following boundary conditions are defined for the the temperature function $\theta^{+}$
\beq
\theta^{+}(x_1, x_2=+\infty)=0, \qquad 
\left.\frac{\partial \theta^{+}}{\partial x_2}\right|_{ x_2=0^+}=-\frac{\theta_s}{2L}\left[\delta\left(\frac{x_1-a_1}{L}\right)-\delta\left(\frac{x_1-a_2}{L}\right)\right].
\label{thetabound}
\eeq

Applying the Fourier transform with respect to the variable $x_1$, the equation \eq{harmapp} becomes
\beq
\tilde{\theta}^{+''}-\xi^2\tilde{\theta}^{+}=0,
\label{harmfou}
\eeq
where $^{'}$ denotes the derivative with respect to $x_2$. The general solution of equation \eq{harmfou} takes the form
\beq
\tilde{\theta}^{+}(\xi, x_2)=C_1e^{-|\xi| x_2}+C_2e^{|\xi| x_2},
\label{solfou}
\eeq
where the constants $C_1$ and $C_2$ are determined by boundary conditions \eq{thetabound}. They are given by
\beq
C_1=\frac{\theta_s}{2|\xi|}(e^{ia_1\xi}-e^{-ia_2\xi}), \qquad C_2=0,
\eeq
consequently, the Fourier transform of the temperature profile for this case becomes
\beq
\tilde{\theta}^{+}(\xi, x_2)=\frac{\theta_s}{2|\xi|}(e^{ia_1\xi}-e^{-ia_2\xi})e^{-|\xi| x_2},
\label{eqfou}
\eeq
and the temperature $\theta^{+}$ is obtained applying the Fourier inversion to expression \eq{eqfou}
\beq
\theta^{+}(x_1, x_2)=-\frac{\theta_s}{4\pi}\left\{\ln\left[(x_1-a_1)^2+x_2^2\right]-\ln\left[(x_1-a_2)^2+x_2^2\right]\right\}.
\label{theta+}
\eeq

Assuming the same profile \eq{qdelta} for the normal heat flux on the boundary $x_2=0^-$, an expression similar to the \eq{theta+} is derived for
the temperature $\theta^{-}$ in the lower half-plane $x_2<0$. The average and the jump of the temperature across the plane $x_2=0$,
which are defined respectively by expressions \eq{average_td}$_{(1)}$ and \eq{jump_td}$_{(1)}$, are then given by 
\beq
\left\langle \theta \right\rangle (x_1)=-\frac{\theta_s}{4\pi}\left[\ln(x_1-a_1)^2-\ln(x_1-a_2)^2\right],\quad \jump{0.1}{\theta}(x_1)=0,
\eeq
and the average and jump of the normal heat flux, defined by expressions \eq{average_flux}$_{(1)}$ and \eq{jump_flux}$_{(1)}$, finally become
\beq
\left\langle q_1 \right\rangle (x_1)=\frac{\theta_s}{4 L}(k_t^+ -k_t^-)\left[\delta\left(\frac{x_1-a_1}{L}\right)-\delta\left(\frac{x_1-a_2}{L}\right)\right],
\eeq
\beq
 \jump{0.1}{q_1}(x_1)=\frac{\theta_s}{2 L}(k_t^+ +k_t^-)\left[\delta\left(\frac{x_1-a_1}{L}\right)-\delta\left(\frac{x_1-a_2}{L}\right)\right].
\eeq

\subsection{Symmetrically distributed temperature profile at the interface}
\label{thetagauss}
The following distributed profile for the temperature on the boundary $x_2=0^{+}$ is assumed
\beq
\theta^{+}(x_1, x_2=0^+)=\frac{\theta_s x_1}{L}e^{-\frac{x_1^2}{L^2}}.
\label{tapp}
\eeq
The boundary conditions for the temperature $\theta^{+}(x_1, x_2)$ are given by expression \eq{tapp}
together with the relation \eq{fields_inf}$_{(1)}$ here reported
\beq
\theta^{+}(x_1, x_2=+\infty)=0.
\label{tapp0}
\eeq
Also in this case, the Fourier transform of the temperature $\tilde{\theta}^{+}(\xi,x_2)$ is given by the solution of equation \eq{harmfou}, and then it 
takes the form \eq{solfou}. The constants $C_1$ and $C_2$, determined by means of boundary conditions \eq{tapp} and \eq{tapp0}, become
\beq
C_1=\frac{i\sqrt{\pi}\theta_s L^2}{2}\xi e^{-\frac{L^2\xi^2}{4}}, \qquad C_2=0,
\eeq
and then the Fourier transform of the temperature profile is given by 
\beq
\tilde{\theta}^{+}(\xi, x_2)=\frac{i\sqrt{\pi}\theta_s L^2}{2}\xi e^{-\left(|\xi| x_2+\frac{L^2\xi^2}{4}\right)}.
\label{thetafou}
\eeq
Applying the inverse Fourier transform to \eq{thetafou}, the following profile for the temperature in the upper half-plane is derived
\begin{displaymath}
 \theta^{+}(x_1, x_2)=\frac{\theta_s}{2}e^{-\frac{x_1^2-x_2^2}{L^2}}\left\{x_1\cos\left(\frac{2x_1x_2}{L^2}\right)+x_2\sin\left(\frac{2x_1x_2}{L^2}\right)\right.
\end{displaymath}
\beq
\left. -x_1\Im\left[e^{-\frac{2ix_1x_2}{L^2}}\mbox{Erfi}\left(\frac{x_1+ix_2}{L}\right)\right]
-x_2\Re\left[e^{-\frac{2ix_1x_2}{L^2}}\mbox{Erfi}\left(\frac{x_1+ix_2}{L}\right)\right]\right\}.
\label{theta+dist}
\eeq

Assuming the same profile \eq{tapp} for the temperature on the boundary $x_2=0^-$, the following expression is derived for
the temperature $\theta^{-}$ in the lower half-plane $x_2<0$:
\begin{displaymath}
 \theta^{-}(x_1, x_2)=\frac{\theta_s}{2}e^{-\frac{x_1^2-x_2^2}{L^2}}\left\{x_1\cos\left(\frac{2x_1x_2}{L^2}\right)+x_2\sin\left(\frac{2x_1x_2}{L^2}\right)\right.
\end{displaymath}
\beq
\left. +x_1\Im\left[e^{-\frac{2ix_1x_2}{L^2}}\mbox{Erfi}\left(\frac{x_1+ix_2}{L}\right)\right]
+x_2\Re\left[e^{-\frac{2ix_1x_2}{L^2}}\mbox{Erfi}\left(\frac{x_1+ix_2}{L}\right)\right]\right\}.
\label{theta-dist}
\eeq
The average and the jump of the temperature across the plane $x_2=0$ containing
both the crack and the interface are then given by 
\beq
\left\langle \theta \right\rangle (x_1)=\frac{\theta_s x_1}{L}e^{-\frac{x_1^2}{L^2}},\quad \jump{0.1}{\theta}(x_1)=0,
\eeq
and the average and the jump of the normal heat flux finally become
\beq
\left\langle q_2 \right\rangle (x_1)=-(k_t^+ -k_t^-)\frac{\theta_s}{L}\left[\left
(\frac{2 x_1^2}{L^2}-1\right)e^{-\frac{x_1^2}{L^2}}\mbox{Erfi}\left(\frac{x_1}{L}\right)-\frac{2 x_1}{L\sqrt{\pi}}\right],
\label{qavapp}
\eeq
\beq
\jump{0.1}{q_1}(x_1)=-2(k_t^+ +k_t^-)\frac{\theta_s}{L}\left[\left
(\frac{2 x_1^2}{L^2}-1\right)e^{-\frac{x_1^2}{L^2}}\mbox{Erfi}\left(\frac{x_1}{L}\right)-\frac{2 x_1}{L\sqrt{\pi}}\right].
\label{qjumpapp}
\eeq
It can be easily verified that flux profiles \eq{qavapp} and \eq{qjumpapp} satisfy the integral balance conditions.
\section{Evaluation of integrals $\mS^{(s-)}$ and $\mS^{(s+)}$ for logarithmic functions}
In this Appendix the explicit derivation of results \eq{s-log} and \eq{s+log} is reported in detail. Assuming that the temperature profile at the interface is given by expression \eq{tlog}
and applying the operator $\mS^{(s-)}$ the following integral expression is obtained
\begin{align}
 \mS^{(s-)}\langle\theta\rangle & =-\frac{\theta_s}{4\pi}\mS^{(s-)}\left\{\ln\left[(x_1-a_1)^2\right] -\ln\left[(x_1-a_2)^2\right]\right\} \nonumber\\
                                & =-\frac{\theta_s}{4\pi^2}\bigg\{\int_{-\infty}^{+\infty}\frac{\ln[(t-a_1)^2]}{x_1-t}dt-
                                \int_{-\infty}^{+\infty}\frac{\ln[(t-a_2)^2]}{x_1-t}dt\bigg\} \nonumber\\
                                & = -\frac{\theta_s}{4\pi^2}[I_1^-(x_1)-I_2^-(x_1)], \qquad x_1<0, \ a_1,a_2>0, \ a_1>a_2.
                                \label{ss-}
\end{align}

In order to evaluate $I_1^-$, a change of variables is operated
\begin{displaymath}
 w=t-a_1, \ \ dw=dt, \ \ x_1-t=h-w, \ \ h=x_1-a_1, \quad \mbox{where} \ a_1>0, \ x_1<0, \ \Rightarrow \ h<0,
\end{displaymath}
and then this term becomes 
\begin{equation}
 I_1^-(x_1)=I_1^-(h)=\mP\int_{-\infty}^{+\infty}\frac{\ln[w^2]}{h-w}dw=\pi^2, \qquad h<0.
\label{I1-}
 \end{equation}
Similarly, for calculating $I_2^-$ the following change of variables is introduced
\begin{displaymath}
 s=t-a_2, \ \ ds=dt, \ \ x_1-t=\ell-s, \ \ \ell=x_1-a_2, \quad \mbox{where} \ a_2>0,\ x_1<0 \ \Rightarrow \  \ell<0,
\end{displaymath}
and this lead to 
\begin{equation}
 I_2^-(x_1)=I_2^-(\ell)=\mP\int_{-\infty}^{+\infty}\frac{\ln[s^2]}{\ell-s}ds=\pi^2, \qquad \ell<0.
\label{I2-}
 \end{equation}
As a consequence, substituting the \eq{I1-} and \eq{I2-} into equation \eq{ss-}, it can be observed that the final result coincides with expression \eq{s-log} introduced in Section \ref{exlog}
\begin{equation}
 \mS^{(s-)}\langle\theta\rangle=-\frac{\theta}{4\pi^2}[I_1^-(h)-I_2^-(\ell)]=0, \qquad \forall h, \ell<0.
\end{equation}

Applying the operator $\mS^{(s+)}$ to the temperature profile \eq{tlog}, an integral expression analogous to \eq{ss-} is derived
\begin{align}
 \mS^{(s+)}\langle\theta\rangle & =-\frac{\theta_s}{4\pi}\mS^{(s+)}\left\{\ln\left[(x_1-a_1)^2\right] -\ln\left[(x_1-a_2)^2\right]\right\} \nonumber\\
                                & =-\frac{\theta_s}{4\pi^2}\bigg\{\int_{-\infty}^{+\infty}\frac{\ln[(t-a_1)^2]}{x_1-t}dt-
                                \int_{-\infty}^{+\infty}\frac{\ln[(t-a_2)^2]}{x_1-t}dt\bigg\} \nonumber\\
                                & = -\frac{\theta_s}{4\pi^2}[I_1^+(x_1)-I_2^+(x_1)], \qquad x_1>0, \ a_1,a_2>0, \ a_1>a_2.
                                \label{ss+}
\end{align}
In order to calculate $I_1^+$, a change of variable identical to that operated for evaluating $I_1^-$ is defined
\begin{displaymath}
 w=t-a_1, \ \ dw=dt, \ \ x_1-t=h-w, \ \ h=x_1-a_1, \quad \mbox{where} \ a_1>0, \ x_1>0.
\end{displaymath}
It is important to note that, differently from the term $I_1^-$, characterized by $h<0$, for the evaluation of $I_1^+$ both the cases $h<0$ and $h>0$ must be considered. Consequently
the result is given by
\begin{equation}
 I_1^+(x_1)=I_1^+(h)=\mP\int_{-\infty}^{+\infty}\frac{\ln[w^2]}{h-w}dw=\left\{
 \begin{array}{cc}
  \pi^2 & h<0, \ \Rightarrow \ x_1<a_1, \\
  -\pi^2 & h>0, \ \Rightarrow \ x_1>a_1,
 \end{array}
\right.
\label{I1+}
\end{equation}
Similarly, for calculating $I_2^+$ the same change of variable used for $I_2^-$ is introduced
\begin{displaymath}
 s=t-a_2, \ \ ds=dt, \ \ x_1-t=\ell-s, \ \ \ell=x_1-a_2, \quad \mbox{where} \ a_2>0,\ x_1>0,
\end{displaymath}
and the result of the integral is given by 
\begin{equation}
 I_2^+(x_1)=I_2^+(\ell)=\mP\int_{-\infty}^{+\infty}\frac{\ln[s^2]}{\ell-s}ds=\left\{
 \begin{array}{cc}
  \pi^2 & \ell<0, \ \Rightarrow \ x_1<a_2, \\
  -\pi^2 & \ell>0, \ \Rightarrow \ x_1>a_2,
 \end{array}
\right.
\label{I2+}
\end{equation}

The sum of the terms \eq{I1+} and \eq{I2+} becomes
\begin{displaymath}
 I_1^+(h)-I_2^+(\ell)=\left\{
 \begin{array}{cl}
 0 & h<0, \ \ell<0, \ \Rightarrow 0<x_1<a_2, \\
 2\pi^2 & h<0, \ \ell>0, \ \Rightarrow a_2<x_1<a_1, \\
 0 & h>0, \ \ell>0, \ \Rightarrow x_1>a_1, 
 \end{array}
 \right.
\end{displaymath}
and then substituting this expression into equation \eq{ss+} the following final result is obtained
\begin{equation}
 \mS^{(s+)}\langle\theta\rangle=\left\{
 \begin{array}{cl}
 0 &   0<x_1<a_2, \\
 -\cfrac{\theta_s}{2} &  a_2<x_1<a_1, \\
 0 &  x_1>a_1, 
 \end{array}
 \right.
 \label{ss+fin}
\end{equation}

The equation \eq{ss+fin} can be expressed by introducing the Heaviside functions $H(a_1-x_1)$ and  $H(a_2-x_1)$ \citep{Arfk1}, 
and then it becomes identical to the \eq{s+log} used in Section \ref{exlog}
\begin{equation}
 \mS^{(s+)}\langle\theta\rangle=-\frac{\theta_s}{2}[H(a_1-x_1)-H(a_2-x_1)],    \qquad a_1>a_2.
\end{equation}
\newpage
\bibliography{SOFC} 
\bibliographystyle{elsarticle-harv}


\end{document}